\documentclass[]{interact}

\usepackage{epstopdf}
\usepackage[caption=false]{subfig}
\usepackage[authoryear]{natbib}

\bibpunct{(}{)}{,}{a}{,}{,}

\renewcommand\bibfont{\fontsize{10}{12}\selectfont}
\usepackage{subcaption}
\theoremstyle{plain}
\newtheorem{theorem}{Theorem}[section]

\theoremstyle{definition}

\theoremstyle{remark}

\usepackage{mathtools}

\usepackage{hyperref} 
\usepackage{algorithm}

\usepackage{algpseudocode}
\usepackage{adjustbox}

\usepackage[left]{lineno}

\usepackage{xcolor}
\usepackage{amsthm}

\begin{document}


\title{Flexible model for varying levels of zeros and  outliers in count data}

\author{
\name{Touqeer Ahmad \textsuperscript{a} and Abid Hussain\textsuperscript{b} \thanks{CONTACT Abid Hussain Email: abid0100@gmail.com}}
\affil{\textsuperscript{a}Department of Mathematics, University of Oslo, P.O. Box 1053, Blindern, Oslo, 0316, Norway. \\ \textsuperscript{b}Department of Statistics, Punjab Higher Education Department, Pakistan.}
}

\maketitle
\date{}
\begin{abstract}
Count regression models are necessary for examining discrete dependent variables alongside covariates. Nonetheless, when data display outliers, overdispersion, and an abundance of zeros, traditional methods like the zero-inflated negative binomial (ZINB) model sometimes do not yield a satisfactory fit, especially in the tail regions. This research presents a versatile, heavy-tailed discrete model as a resilient substitute for the ZINB model. The suggested framework is built by extending the generalized Pareto distribution and its zero-inflated version to the discrete domain. This formulation efficiently addresses both overdispersion and zero inflation, providing increased flexibility for heavy-tailed count data. Through intensive simulation studies and real-world implementations, the proposed models are thoroughly tested to see how well they work. The results show that our models always do better than classic negative binomial and zero-inflated negative binomial regressions when it comes to goodness-of-fit. This is especially true for datasets with a lot of zeros and outliers. These results highlight the proposed framework's potential as a strong and flexible option for modeling complicated count data.
\end{abstract}


\section{Introduction}\label{sec:1}
\begin{figure}[!t]
\centering
\subfloat[Low level outliers]{%
\resizebox*{6cm}{!}{\includegraphics{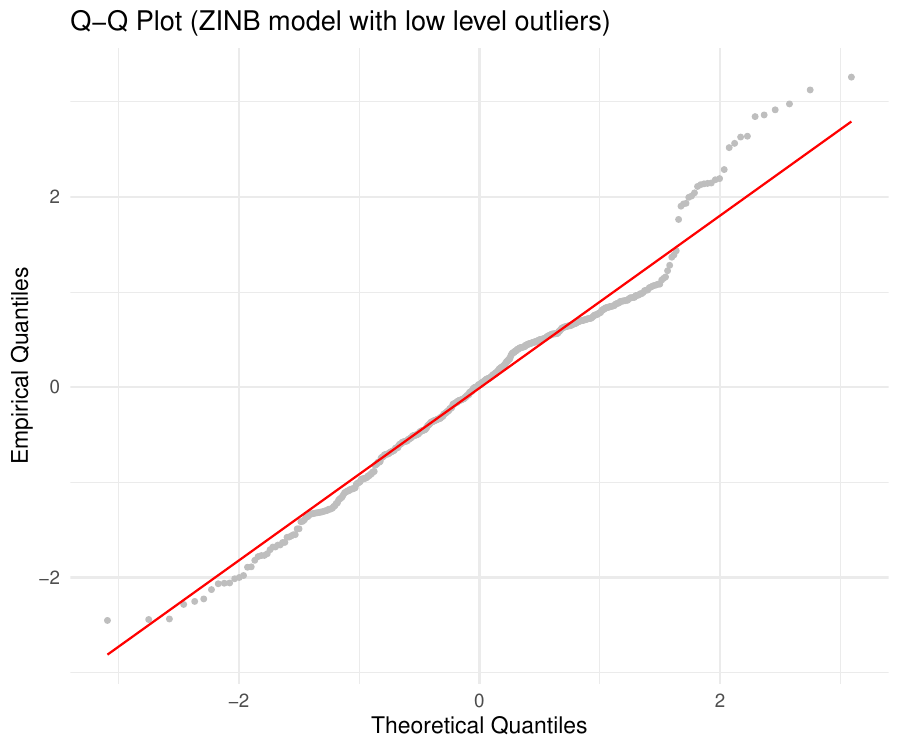}}}\hspace{5pt}
\subfloat[Medium level outliers]{%
\resizebox*{6cm}{!}{\includegraphics{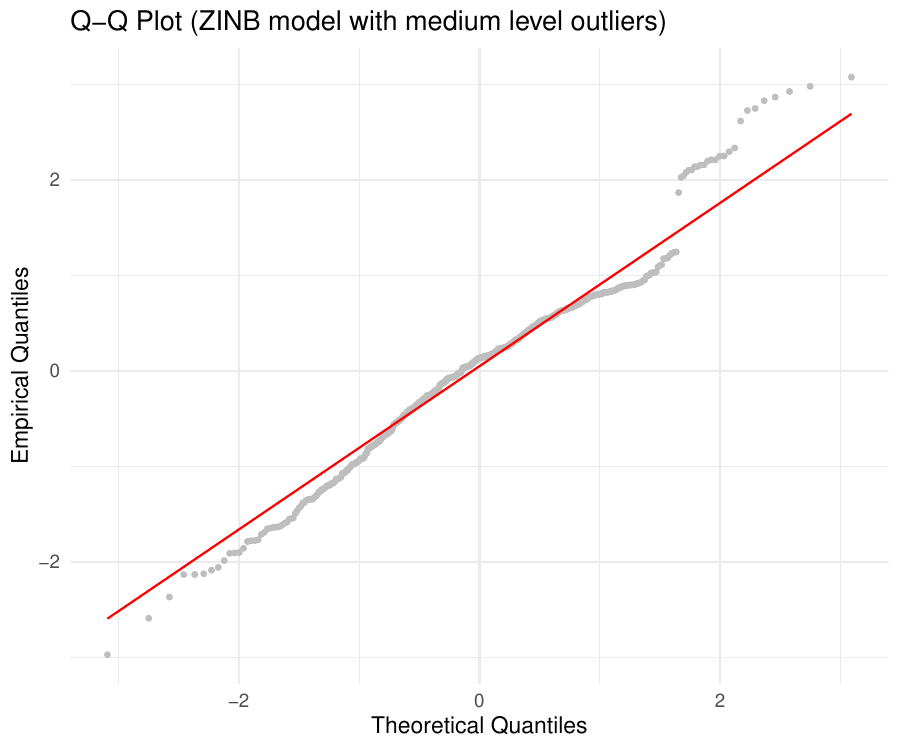}}}
\subfloat[High level outliers]{%
\resizebox*{6cm}{!}{\includegraphics{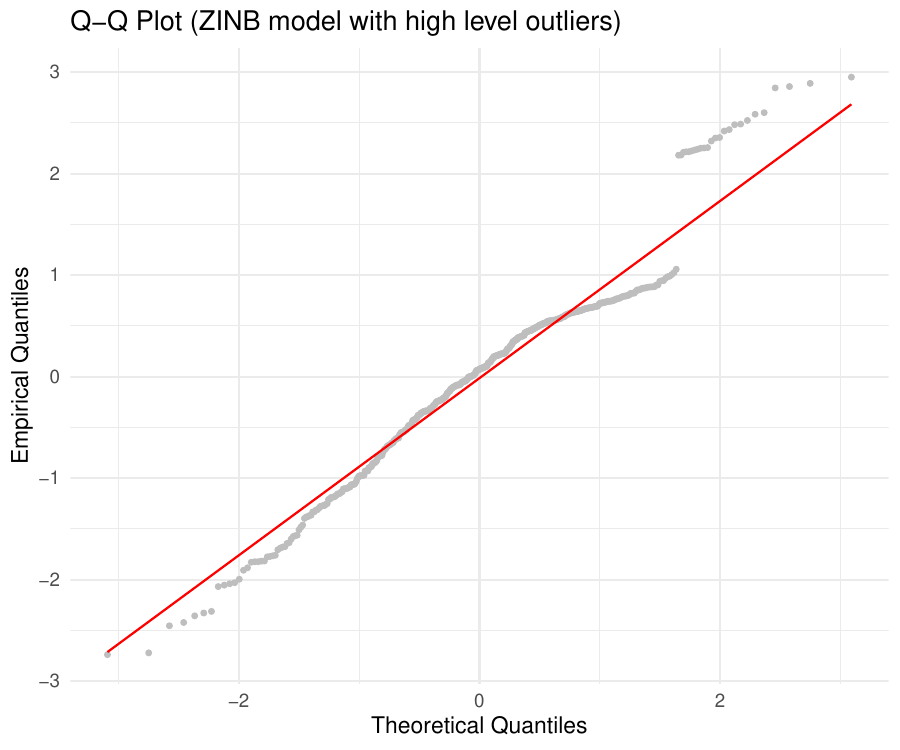}}}
\caption{Randomized residual Q-Q plots of fitted ZINBD regression to the count data sets simulated by following the steps given in \citet{tuzen2020simulation}.}  \label{fig:existmodel}
\end{figure}

The statistical literature has various studies focused on the modeling of zero-inflated data affected by outliers. In a sample research, \cite{tuzen2020simulation} produced data from a negative binomial distribution, encompassing varying zero proportions (25\%, 50\%, and 70\%) coupled with distinct outlier levels—specifically, low (1\%), medium (5\%), and high (10\%). Their findings demonstrated that the zero-inflated negative binomial (ZINB) regression model exhibits strong performance according to the Akaike information criterion (AIC), notwithstanding its inadequacy in delivering an acceptable graphical fit within the outlier region. ZINB works well for modeling data that is overdispersed, but it doesn't work well for dealing with outliers. Following \cite{tuzen2020simulation}, we simulate $n=500$ data points from a Negative Binomial Distribution (NBD) with an additional 50\% of zeros and varying levels of outliers in the response variable, and then fit a ZINB regression with covariates $X_1$ and $X_2$; details regarding simulation of these covariates are provided in Section~\ref{sec:simulation}. Figure \ref{fig:existmodel} illustrates that the ZINB regression model adequately captures the bulk of the distribution but fails to model the tail where outliers reside. In order to address these limitations, a model should represent both the bulk and the tail of the distribution accurately, thus addressing the limitations of ZINB regression when there are outliers.

Zero-inflation is a common phenomenon in count data, occurring when the number of zeros exceeds what standard distributions such as Poisson or NB can accommodate. This happens frequently in many real-life situations, like insurance, healthcare, and epidemiology. For example, policyholders may not file any claims, doctors may not see any patients, or there may not be any new COVID-19 cases each day. The creation of zero-inflated alternatives to traditional count models, such as the Zero-Inflated Poisson (ZIP) and Zero-Inflated Negative Binomial (ZINB) regressions, gives a clear solution to this issue.  Their integration into standard statistical software has enabled extensive utilization in applied research. This is substantiated by \cite{avci2015comparison} and \citet{cahoy2021flexible}, they assessed various over-dispersed models, and \cite{ismail2013estimation}, which used ZIP and ZINB regressions on Malaysian own-damage claim data. Further demonstrations of their efficacy include applications to crash data \citep{lord2005poisson} and urban highway incidents \citep{ayati2014modeling}.

Recent studies have enhanced the modeling of zero-inflated count data. For example, \citet{LIM2014151} advanced the modeling of count data by combining zero-inflated Poisson and finite mixture approaches to effectively address excess zeros and unobserved heterogeneity. \citet{li2021bayesian} 
introduce a Bayesian hierarchical model for spatial transcriptomics data that effectively handles zero inflation and overdispersion using a zero-inflated negative binomial framework.
Additionally, \citet{altinisik2023addressing} propose novel multilevel heterogeneous hurdle models to jointly address overdispersion, zero-inflation, and unobserved heterogeneity in clustered count data. By incorporating Poisson–Lindley and Poisson–Ailamujia distributions, the study enhances modeling flexibility and predictive performance for complex hierarchical datasets.
\citet{sidumo2024count} explores overdispersion in ecological count data by comparing traditional count regression models with machine learning techniques. Their findings highlight the superior predictive performance of ML methods, encouraging their broader application in ecological modeling.

The issue of outliers in count data has simultaneously been a topic of considerable examination. These findings, which are numerically remote from the majority of the sample, can significantly undermine the validity of conventional statistical methods. A significant risk lies in the susceptibility of the sample mean ($\bar{x}$), which might be unduly affected by an isolated extreme result. This issue is intensified in zero-inflated models, such as the ZIP distribution, where a prevalence of zeros is typical \citep{yang2011outlier}. The identification of outliers is somewhat subjective due to the lack of a strict mathematical definition; however, their negative effects are well-documented. Thus, traditional outlier detection techniques (e.g., Chauvenet’s criterion, Grubbs’ test) are inadequate for asymmetric, non-normal count distributions such as the ZIP.

The literature on robust modeling for count data with outliers is expanding. \cite{ascari2021new} introduced a robust regression framework for overdispersed count data, demonstrating its capability to address both outlier contamination and zero-inflation simultaneously. Similarly, \citet{chen2021robust} developed a robust negative binomial regression based on a weighted maximum likelihood approach, which downweights the influence of outliers during estimation. More recently, \citet{bourguignon2025general} explored robust beta regression for rates and proportions, reflecting a broader trend toward models that are inherently resistant to anomalous observations. However, many of these robust methods are not explicitly designed to handle the joint challenges of zero-inflation and extreme values in the upper tail.

Motivated by the need to model zero-inflated, over-dispersed, and outlier-prone count data, this study compares NB and ZINB models with a new flexible model that employs a regularly varying tail approximation via the tail index of the generalized Pareto distribution. We refer to these models as the discrete extended generalized Pareto distribution (DEGPD) and its zero-inflated counterpart (ZIDEGPD). Originally proposed by \citet{ahmad2025extended} for modeling nonlinear effects of environmental covariates in avalanche counts, these models offer three key advantages: (1) they accurately represent the full range of non-negative integer-valued data, making them suitable for diverse count data applications, with particular flexibility in modeling tail events and outliers; (2) they deliver reliable performance on zero-inflated datasets, irrespective of the proportion of zeros; and (3) they provide a robust and flexible framework for discrete data, especially in scenarios where conventional models are inadequate for handling outliers (see Figure \ref{fig:existmodel}).

The remainder of this paper is organized as follows. Section~\ref{MF} provides an overview of conventional count regression models, including NB and ZINB, and introduces the heavy-tailed count regression models based on DEGPD and ZIDEGPD within a unified distributional regression framework. Section~\ref{sec:simulation} presents a comprehensive simulation study, and Section~\ref{sec:realapp} applies the count regression models to Global Terrorism Database (GTD)\footnote{Acknowledged by authors: START (National Consortium for the Study of Terrorism and Responses to Terrorism). (2022). Global Terrorism Database 1970 - 2020 [data file]. https://www.start.umd.edu/gtd} recorded fatality data from Afghanistan. Finally, Section~\ref{sec:conl} concludes the paper.

\section{Count data regression models}\label{MF}

\subsection{Negative binomial model}\label{sec:negb}
The NB regression model is a flexible extension of the Poisson model that accounts for overdispersion, where the variance exceeds the mean. This overdispersion often arises from unobserved heterogeneity among observations. The NB distribution can be derived as a Poisson-Gamma mixture and is commonly parameterized in the NB2 form.

Let $Z$ be a count random variable with conditional mean $\mu>0$ and dispersion parameter $\alpha > 0$. The probability mass function (PMF) is given by
\begin{equation}
\Pr(Z = k \mid \mu, \alpha) = \frac{\Gamma(k + \alpha^{-1})}{\Gamma(k+1)\Gamma(\alpha^{-1})} \left( \frac{\alpha^{-1}}{\alpha^{-1} + \mu} \right)^{\alpha^{-1}} \left( \frac{\mu}{\alpha^{-1} + \mu} \right)^k, \quad k \in \mathbb{N}_0,
\label{eq:nb_pmf}
\end{equation}
where $\Gamma(\cdot)$ denotes the Gamma function. The variance of $Z$ is $\text{Var}(Z \mid \mathbf{x}) = \mu + \alpha \mu^2$, indicating that the variance exceeds the mean for any $\alpha > 0$. As $\alpha \to 0$, the NB model converges to the Poisson model. To link the NB model with the distributional framework described in Section~\ref{distregression}, the log-likelihood function for (\ref{eq:nb_pmf}) is defined as
\begin{equation}
\mathcal{L}(\boldsymbol{\psi}) = \sum_i \log \Pr(Z_i = k_i \mid \mu_i, \alpha_i).
\label{eq:nb_loglik}
\end{equation}
The full form of log-likelihood expression of the NB model is provided in equation~\eqref{eq:app_nb_loglik} of Appendix~\ref{lik-appendix}.

\subsection{Zero-inflated negative binomial model}\label{sec:zinegb}
Count data often exhibit an excess of zeros beyond what the standard NB model predicts. The ZINB model addresses this by incorporating an additional process for structural zeros. Specifically, zeros can arise from a ``certain zero" process or from the standard NB count process.

Let $Z^*$ denote the observed random variable, and let $0 \leq \omega \leq 1$ represent the probability of a structural zero. The PMF of the ZINB model is defined as
\begin{equation}
\Pr(Z^* = k^*) = 
\begin{cases}
\omega + (1 - \omega) \Pr(Z = 0 \mid \mu, \alpha), & k^* = 0, \\
(1 - \omega) \Pr(Z = k^* \mid \mu, \alpha), & k^* > 0,
\end{cases}
\label{eq:zinb_pmf}
\end{equation}
where $\Pr(Z = k^* \mid \mu, \alpha)$ follows the NB PMF given in (\ref{eq:nb_pmf}). The cumulative distribution function (CDF) is given by
\begin{equation}
\Pr(Z^* \leq k^*) = \omega + (1 - \omega) F_{\text{NB}}(k^*; \mu, \alpha),
\label{eq:zinb_cdf}
\end{equation}
where $F_{\text{NB}}(\cdot; \mu, \alpha)$ denotes the CDF of the NB distribution. The log-likelihood function for (\ref{eq:zinb_pmf}) is
\begin{equation}
\mathcal{L}(\boldsymbol{\psi}) = \sum_i \log \Pr(Z^*_i = k^*_i \mid \mu_i, \alpha_i, \omega_i).
\label{eq:zinb_loglik}
\end{equation}
The log-likelihood expression of the ZINB model is given in equation~\eqref{eq:app_zinb_loglik} of Appendix~\ref{lik-appendix}.

\subsection{Heavy-tailed extreme value model}\label{sec:degpd}
Our technique is fundamentally based on a significant extension of the generalized Pareto distribution (GPD) introduced by \citet{papastathopoulos2013extended}, which includes an additional shape parameter. This adjustment enhances the model's adaptability while maintaining its asymptotic upper-tail properties, hence enabling more accurate threshold estimation and permitting the application of lower thresholds. Building upon this study, \citet{naveau2016modeling} developed a theoretical framework that concurrently models the lower tail and the core segment of a distribution, adhering to the tenets of extreme value theory (EVT). Consequently, they specified essential criteria for EVT-consistency and developed a technique for smooth interpolation between the lower and upper bounds over the unit interval $[0,1]$.

Utilizing this methodology, \citet{ahmad2025extended} developed a model appropriate for discrete data throughout its entire support. Their approach utilizes an integral transformation to simulate random variables from a Generalized Pareto Distribution (GPD) through the inverse transform method: $\mathcal{H}^{-1}_{\sigma,\gamma}(U)$, where $U \sim \mathcal{U}(0,1)$ represents a uniform random variable. A new random variable is defined by extending this concept as follows
\begin{equation}
X = \mathcal{H}^{-1}_{\sigma,\gamma} \left( \mathcal{R}^{-1}(U) \right),
\label{eq:x_transform}
\end{equation}
where $\mathcal{R}$ is a cumulative distribution function (CDF) with support on $[0,1]$. The resulting distribution of $X$ has the CDF
\begin{equation}
\overline{\mathcal{H}}_{\sigma,\gamma,\alpha}(x) = \mathcal{R}(\mathcal{H}(x; \sigma, \gamma), \alpha).
\label{eq:egpd_cdf}
\end{equation}
A primary problem becomes identifying a suitable function $\mathcal{R}$ that maintains the upper-tail characteristics dictated by the shape parameter $\gamma$, especially in the context of outliers, while concurrently allowing for precise modeling of the lower tail. \citet{naveau2016modeling} introduced specific criteria to ensure the validity of such transformations.

Let $Z \in \mathbb{N}_0$ be a discrete random variable. We consider the CDF $\mathcal{R}(u; \alpha) = u^\alpha$, $\alpha > 0$, which satisfies the conditions in \citet{naveau2016modeling}. We derive a discrete distribution by discretizing the CDF (\ref{eq:egpd_cdf}) as
\begin{equation}
\begin{aligned}
\Pr(Z = k) &= \overline{\mathcal{H}}_{\sigma,\gamma,\alpha}(k+1) - \overline{\mathcal{H}}_{\sigma,\gamma,\alpha}(k), \quad k \in \mathbb{N}_0, \\
&= \left[1 - \left(1 + \frac{(k+1)\gamma}{\sigma} \right)^{-\frac{1}{\gamma}} \right]^\alpha - \left[1 - \left(1 + \frac{k\gamma}{\sigma} \right)^{-\frac{1}{\gamma}} \right]^\alpha.
\end{aligned}
\label{eq:degpd_pmf}
\end{equation}
We refer to (\ref{eq:degpd_pmf}) as the discrete extended generalized Pareto distribution (DEGPD). The CDF of DEGPD is
\begin{equation}
\Pr(Z \leq k) = \overline{\mathcal{H}}_{\sigma,\gamma,\alpha}(k+1) = \mathcal{R} \left( \mathcal{H}(k+1; \sigma, \gamma), \alpha \right) = \left[1 - \left(1 + \frac{(k+1)\gamma}{\sigma} \right)^{-\frac{1}{\gamma}} \right]^\alpha,
\label{eq:degpd_cdf}
\end{equation}
and the inverse CDF is
\begin{equation}
Q_p = 
\begin{cases}
\left\lceil \frac{\sigma}{\gamma} \left\{ \left(1 - p^{1/\alpha} \right)^{-\gamma} - 1 \right\} \right\rceil - 1, & \text{if } \gamma > 0, \\
\left\lceil -\sigma \log \left(1 - p^{1/\alpha} \right) \right\rceil - 1, & \text{if } \gamma = 0.
\end{cases}
\label{eq:degpd_invcdf}
\end{equation}

For the model in (\ref{eq:degpd_pmf}), we say that a non-negative discrete random variable $Z$ belongs to the \emph{discrete maximum domain of attraction}, denoted $Z \in \mathrm{d\text{-}MDA}_\gamma$, if there exists a continuous random variable $X \in \mathrm{MDA}_\gamma$ with shape parameter $\gamma \geq 0$ such that $\Pr(Z \geq k) = \Pr(X \geq k)$ for $k \in \mathbb{N}_0$. That is, $Z$ and the integer part of $X$ share the same distribution, written $Z \stackrel{d}{=} \lfloor X \rfloor$. The variable $X$ is an extension of $Z$, and such an extension is generally not unique \citep{hitz_davis_samorodnitsky_2024}.

Furthermore, \citet{shimura2012discretization} established that a discrete random variable $Z$ belongs to $\mathrm{MDA}_\gamma$ for some $\gamma \geq 0$ if and only if $Z \in \mathrm{d\text{-}MDA}_\gamma$ and $Z$ has a long-tailed distribution. They also showed that several standard discrete distributions—geometric, Poisson, and negative binomial—belong to the class $\mathrm{d\text{-}MDA}$.
\begin{theorem}\label{theorem1}
Let $X$ follow a distribution having a CDF given in (\ref{eq:egpd_cdf}) with shape parameter $\gamma > 0$, and define the discrete random variable $Z = \lfloor X \rfloor$. Then, for each integer $k \in \mathbb{N}_0$, the discrete survival function $Z$ is given by
\[
S_Z(k) := \Pr(Z \geq k) = \Pr(X \geq k) = S_X(k).
\]
Moreover, as $k \to \infty$,
\[
S_Z(k) \sim \alpha \left( \frac{\sigma}{\gamma} \right)^{1/\gamma} k^{-1/\gamma}.
\]
Hence, $S_Z \in RV_{-1/\gamma}$; that is, the DEGPD retains the same tail index as its continuous counterpart. Consequently, the distribution function $F_Z$ given in (\ref{eq:degpd_cdf}) belongs to the discrete maximum domain of attraction with index $\gamma$.
\label{thm:tail}
\end{theorem}
The proof of Theorem \ref{thm:tail} is provided in the Appendix~\ref{theorem-proof}.

\subsection{Zero-inflated heavy-tailed extreme value model}\label{sec:zidegpd}
Suppose the observed random variable $Z^*$ exhibits an excess of zeros compared to the distribution defined in (\ref{eq:degpd_pmf}). To account for this zero inflation, we introduce an additional parameter $0 \leq \omega \leq 1$, representing the probability mass at zero. The resulting PMF is
\begin{equation}
\Pr(Z^* = k^*) = 
\begin{cases}
\omega + (1 - \omega) \overline{\mathcal{H}}_{\sigma,\gamma,\alpha}(1), & k^* = 0, \\
(1 - \omega) \left[ \overline{\mathcal{H}}_{\sigma,\gamma,\alpha}(k^*+1) - \overline{\mathcal{H}}_{\sigma,\gamma,\alpha}(k^*) \right], & k^* = 1, 2, \ldots,
\end{cases}
\label{eq:zidegpd_pmf}
\end{equation}
which can be written explicitly as
\begin{equation}
\Pr(Z^* = k^*) = 
\begin{cases}
\omega + (1 - \omega) \left[ 1 - \left(1 + \frac{\gamma}{\sigma} \right)_+^{-1/\gamma} \right]^\alpha, & \\
(1 - \omega) \left[ \left( 1 - \left(1 + \frac{(k^*+1)\gamma}{\sigma} \right)_+^{-1/\gamma} \right)^\alpha - \left( 1 - \left(1 + \frac{k^*\gamma}{\sigma} \right)_+^{-1/\gamma} \right)^\alpha \right].
\end{cases}
\end{equation}
Expression (\ref{eq:zidegpd_pmf}) is referred to as the zero-inflated DEGPD (ZIDEGPD). The CDF of ZIDEGPD is
\begin{equation}
\Pr(Z^* \leq k^*) = \omega + (1 - \omega) \overline{\mathcal{H}}_{\sigma,\gamma,\alpha}(k^* + 1) = \omega + (1 - \omega) \left[ 1 - \left(1 + \frac{(k^*+1)\gamma}{\sigma} \right)^{-1/\gamma} \right]^\alpha,
\label{eq:zidegpd_cdf}
\end{equation}
and the inverse CDF is
\begin{equation}
Q_{p^*} = 
\begin{cases}
\left\lceil \frac{\sigma}{\gamma} \left[ \left( 1 - (p^*)^{1/\alpha} \right)^{-\gamma} - 1 \right] \right\rceil - 1, & \gamma > 0, \\
\left\lceil -\sigma \log \left( 1 - (p^*)^{1/\alpha} \right) \right\rceil - 1, & \gamma = 0,
\end{cases}
\label{eq:zidegpd_invcdf}
\end{equation}
where $0 < p < 1$ and $0 < p^* = (p - \omega)/(1 - \omega) < 1$. For further details see \cite{ahmad2025extended,ahmad2024new}. The log-likelihood for DEGPD is
\begin{equation}
\mathcal{L}(\boldsymbol{\psi}) = \sum_i \log \Pr(Z_i = k_i \mid \sigma_i, \gamma_i, \alpha_i),
\label{eq:degpd_loglik}
\end{equation}
and for ZIDEGPD
\begin{equation}
\mathcal{L}(\boldsymbol{\psi}) = \sum_i \log \Pr(Z^*_i = k^*_i \mid \sigma_i, \gamma_i, \alpha_i, \omega_i).
\label{eq:zidegpd_loglik}
\end{equation}
The complete likelihood expressions are provided in expressions~\eqref{eq:app_degpd_loglik} and~\eqref{eq:app_zidegpd_loglik} of Appendix~\ref{lik-appendix}.

\subsection{Distribution-based regression framework}\label{distregression}
In the unified distributional regression framework, the parameters of all considered models (NB, ZINB, DEGPD, ZIDEGPD) are expressed as functions of covariates $\mathbf{x} = (x_1, \ldots, x_p)^\top$:
\[
\boldsymbol{\psi}(\mathbf{x}) = (\psi_1(\mathbf{x}), \ldots, \psi_d(\mathbf{x}))^\top.
\]
Each distributional parameter $\psi_i(\mathbf{x})$ is linked to a structured additive predictor
\[
\eta_i(\mathbf{x}) = f_{i1}(\mathbf{x}) + \cdots + f_{iJ_i}(\mathbf{x}),
\]
where the functions $f_{ij}(\cdot)$ may include linear or smooth components. These predictors are connected to their parameters via link functions, ensuring positivity or mapping to the unit interval as required. For example, in NB regression, the parameter vector is $\boldsymbol{\psi}(\mathbf{x}) = (\mu(\mathbf{x}), \alpha(\mathbf{x}))^\top$ with $\mu(\mathbf{x}) = \exp(\eta_\mu(\mathbf{x}))$ and $\alpha(\mathbf{x}) = \exp(\eta_\alpha(\mathbf{x}))$. For the DEGPD model, we use the following link functions
\[
\sigma(\mathbf{x}) = \exp(\eta_\sigma(\mathbf{x})), \quad \gamma(\mathbf{x}) = \exp(\eta_\gamma(\mathbf{x})), \quad \alpha(\mathbf{x}) = \exp(\eta_\alpha(\mathbf{x})),
\]
where log-links ensure positivity. For the zero-inflated extensions (ZIDEGPD and ZINB), a logit link is applied to the zero-inflation probability
\[
\omega(\mathbf{x}) = \frac{\exp(\eta_\omega(\mathbf{x}))}{1 + \exp(\eta_\omega(\mathbf{x}))}.
\]
The additive terms $f_{ij}(\cdot)$ are represented using basis expansions
\[
f_{ij}(x) = \sum_{k=1}^{K_{ij}} \beta_{ijk} B_k(x),
\]
where $B_k(x)$ are known basis functions (e.g., spline bases), and $\beta_{ijk}$ are coefficients to be estimated. Model fitting is performed by maximizing the log-likelihood function $\mathcal{L}(\boldsymbol{\psi})$. Although not required for our specific case study, the proposed implementation can accommodate non-parametric components. Smoothness is enforced through quadratic penalties, leading to the penalized log-likelihood function:
\begin{equation} 
\ell_p = \mathcal{L}(\boldsymbol{\psi}) - \frac{1}{2} \sum_{i=1}^d \sum_{j=1}^{J_i} \lambda_{ij} \boldsymbol{\beta}_{ij}^\top \mathbf{W}_{ij} \boldsymbol{\beta}_{ij}, 
\label{eq:penalized_likelihood} 
\end{equation}
where $\mathcal{L}(\boldsymbol{\psi})$ denotes the log-likelihood of the respective distribution (NB, ZINB, DEGPD, or ZIDEGPD). This cohesive framework enables the estimation of all models through a consistent GAM-based methodology, hence ensuring comparability among various count data models. This study focuses solely on linear specifications; nonetheless, the system inherently adapts to incorporate nonlinear setups.

\begin{figure}
\centering
\includegraphics[width=0.7\linewidth]{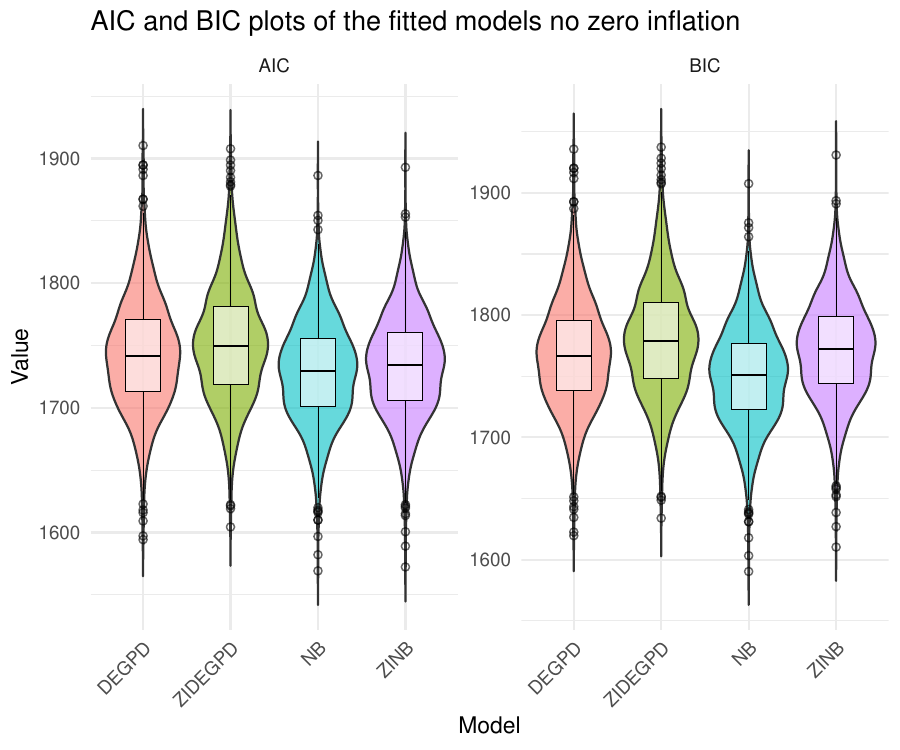}
\caption{Violin boxplots of AIC and BIC over $1000$ replications of the fitted models to data simulated from the negative binomial distribution without zero inflation and outliers.  }
\label{fig:no-zero-inflation}
\end{figure}
\section{Simulation study}
\label{sec:simulation}
To thoroughly assess the performance of the proposed DEGPD and its zero-inflated variant ZIDEGPD, we executed a comprehensive simulation analysis. The main aim was to evaluate their resilience and fitting precision under two significant data complexities—excess zeros and outliers—by comparing them to the established standards of the NB and ZINB models.

Synthetic datasets were generated by manipulating three essential variables inside a controlled framework: (1) The proportion of structural zeros was established at 0\%, 25\%, 50\%, and 70\%; (2) the percentage of observations substituted with extreme values was determined at 0\%, 1\%, 5\%, and 10\%; and (3) when applicable, outliers were produced from three uniform distributions indicative of varying extremity levels—low ($U(30,49)$), medium ($U(50,69)$), and high ($U(70,89)$). With this full-factorial approach, we can study both the individual and interacting impacts of zero-inflation and outliers on model performance.

For each of the $R=1000$ simulation repetitions, we produced a dataset of size $n=500$ in accordance with Algorithm~\ref{alg:data_gen}. The data generation technique guaranteed that the data exhibited the overdispersion characteristic of the NBD, while being systematically tainted by excess zeros and extreme values. We fitted the NB, ZINB, DEGPD, and ZIDEGPD regression models to each simulated dataset. The evaluation of model performance was conducted utilizing the Akaike Information Criterion (AIC) and the Bayesian Information Criterion (BIC), computed as
\[
\text{AIC} = -2\log(\mathcal{L}) + 2p \quad \text{and} \quad \text{BIC} = -2\log(\mathcal{L}) + p\log(n),
\]
where $\mathcal{L}$ is the maximized likelihood and $p$ is the number of parameters. Lower AIC/BIC values indicate a better balance between model fit and complexity.

\begin{algorithm}[!t]
\caption{Data generation procedure}\label{alg:data_gen}
\begin{algorithmic}[1]
\Require Sample size $n=500$, zero-inflation rate $p_z$, outlier ratio $p_o$, outlier range $[a, b]$.
\Ensure Simulated dataset $\{(Y_{i},X_{1i},X_{2i})\}_{i=1}^{n}$.
\State Set the number of outlier observations: $n_{o} = \lfloor n \times p_o \rfloor$
\State Set the number of regular observations: $n_{1} = n - n_{o}$
\State Generate covariates:
\State \quad $X_{1} \sim \text{Categorical}(0.5, 0.5)$ for levels $\{1,2\}$.
\State \quad $X_{2} \sim \text{Categorical}(0.40, 0.35, 0.25)$ for levels $\{1,2,3\}$.
\State Compute the mean structure for each observation $i$:
\[
\log(\mu_{i})=\beta_{0} + \beta_{1}X_{1i} + \beta_{2}X_{2i}, \quad \text{with } \bm{\beta} = (1, 0.5, -0.7)^{T}.
\]
\State Generate $n_1$ base counts: $Y^{\text{base}}_i \sim \text{NB}(\mu_{i}, \phi=2)$.
\State Introduce zero-inflation: Randomly set $n \times p_z$ of the $Y^{\text{base}}_i$ values to zero.
\State Generate $n_{o}$ outlier values: $Y^{\text{outlier}}_j \sim U(a, b)$ for $j=1, \ldots, n_o$.
\State Replace a randomly selected subset of $n_o$ observations in $Y^{\text{base}}$ with the $Y^{\text{outlier}}$ values.
\State Combine to form the final response vector $Y$.
\end{algorithmic}
\end{algorithm}

\begin{figure}
\centering
\includegraphics[width=4.5cm,height=5.5cm]{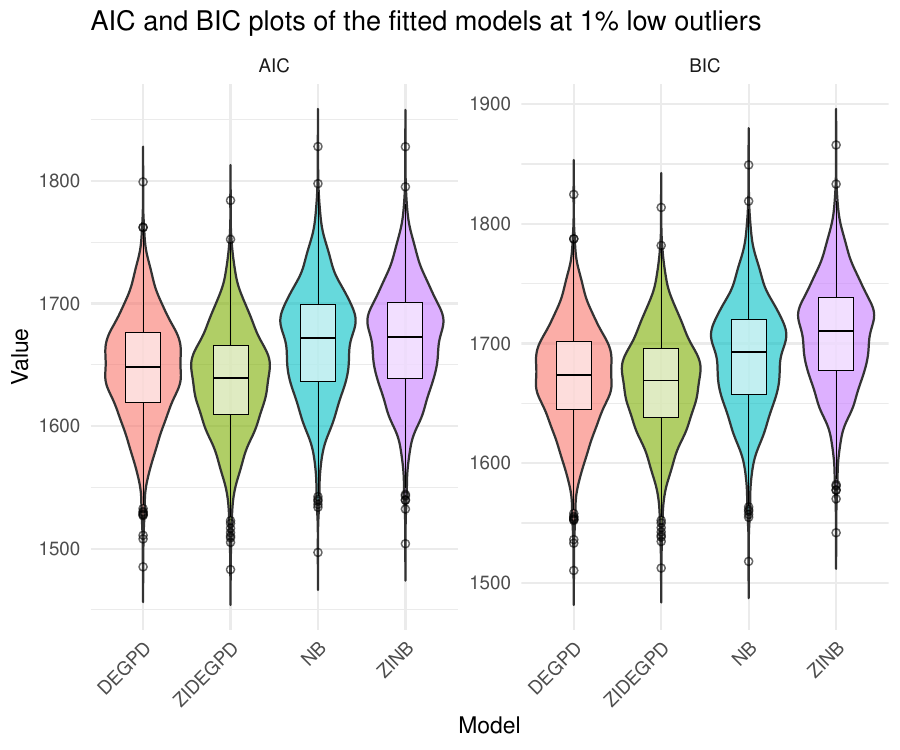}
\includegraphics[width=4.5cm,height=5.5cm]{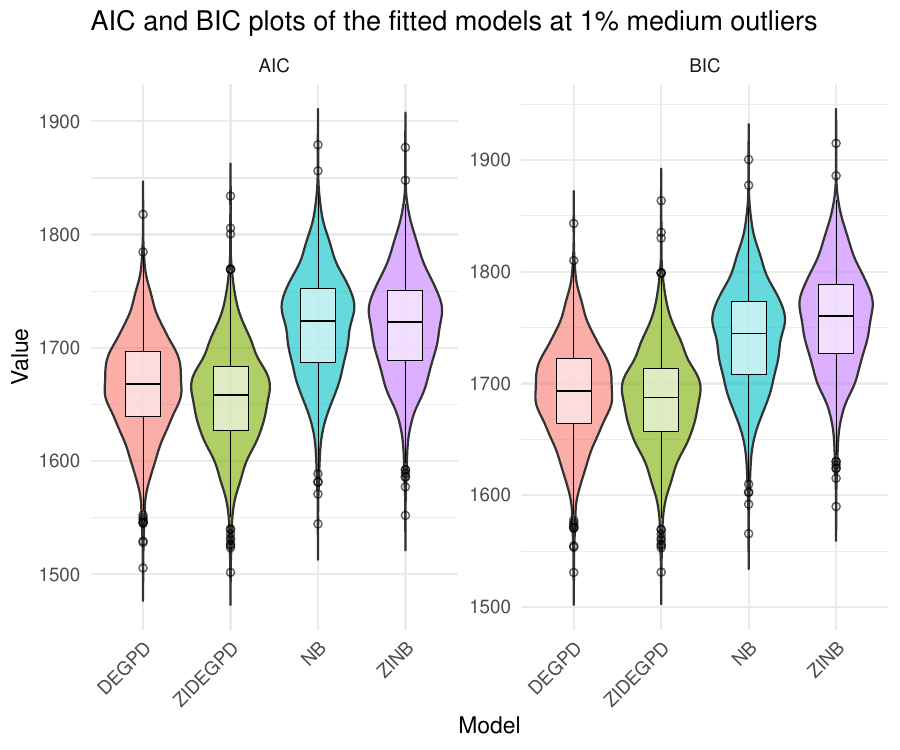}
\includegraphics[width=4.5cm,height=5.5cm]{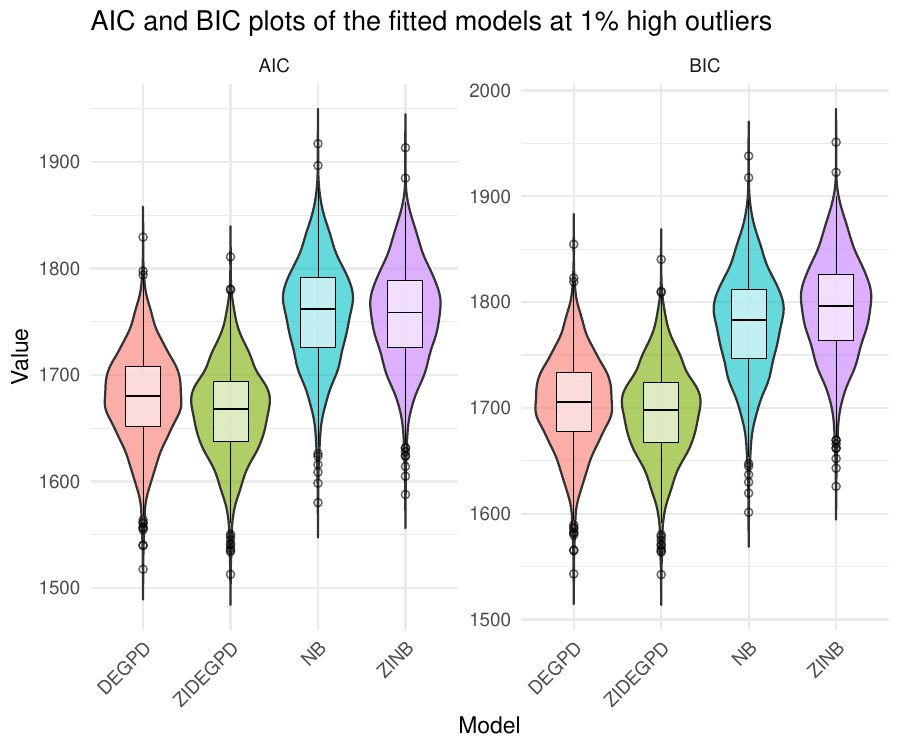}
\includegraphics[width=4.5cm,height=5.5cm]{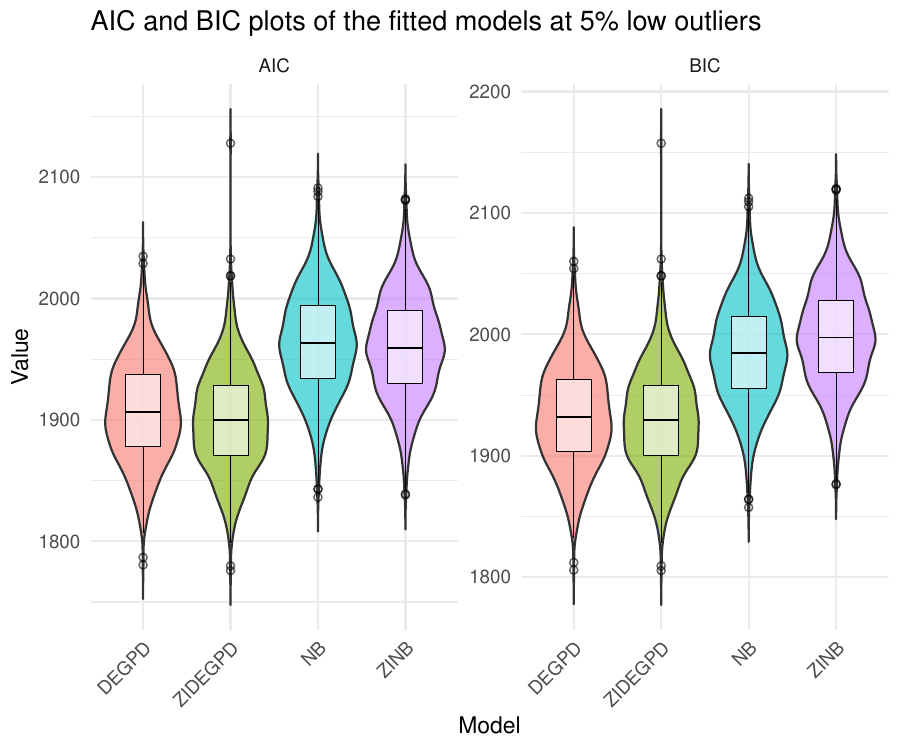}
\includegraphics[width=4.5cm,height=5.5cm]{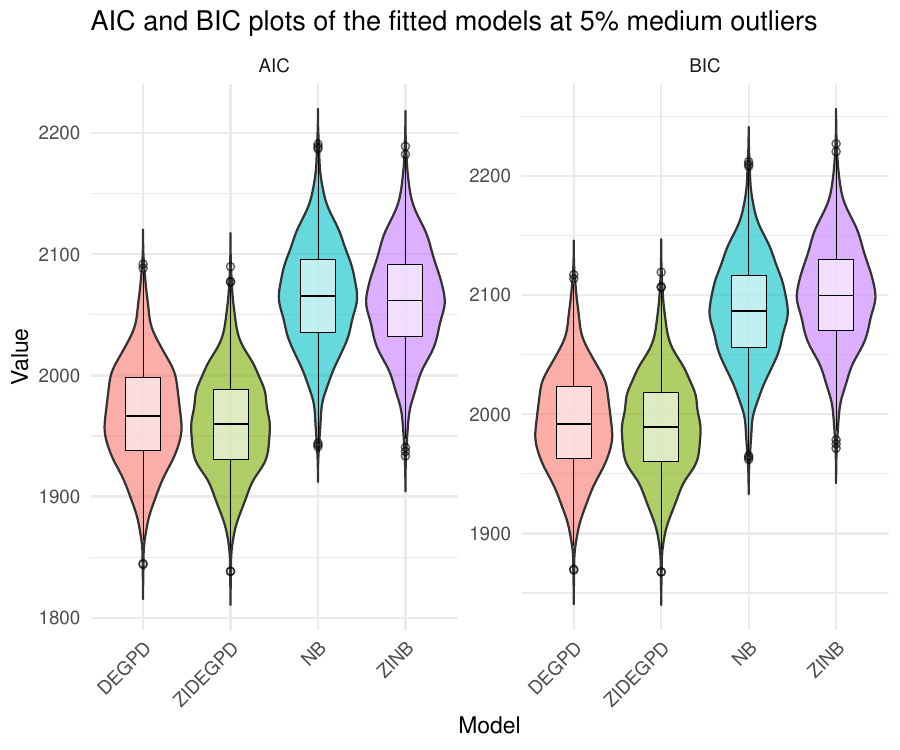}
\includegraphics[width=4.5cm,height=5.5cm]{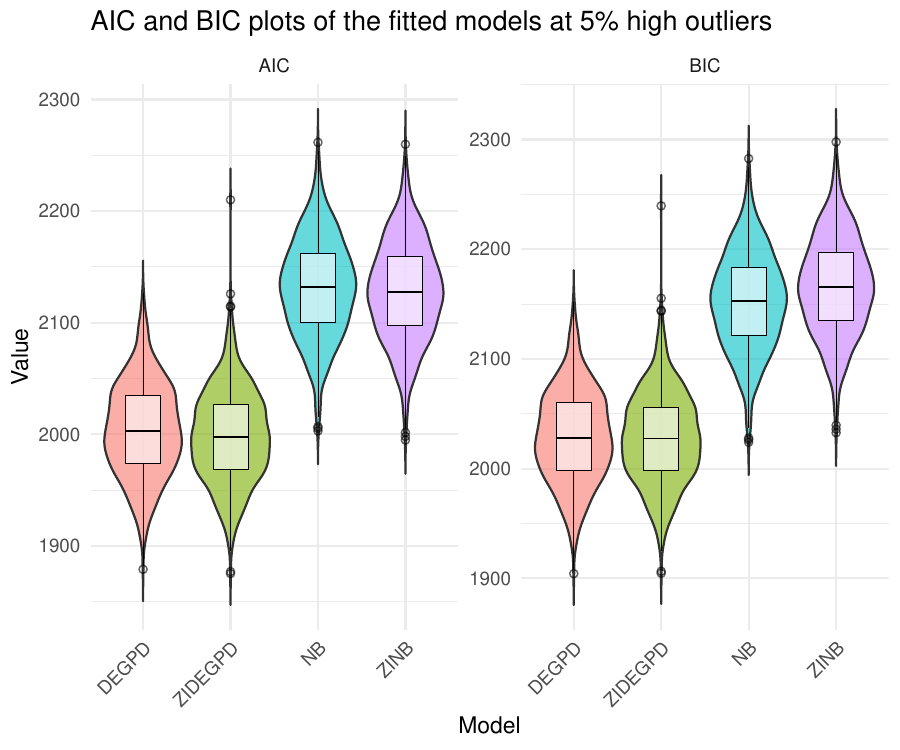}
\includegraphics[width=4.5cm,height=5.5cm]{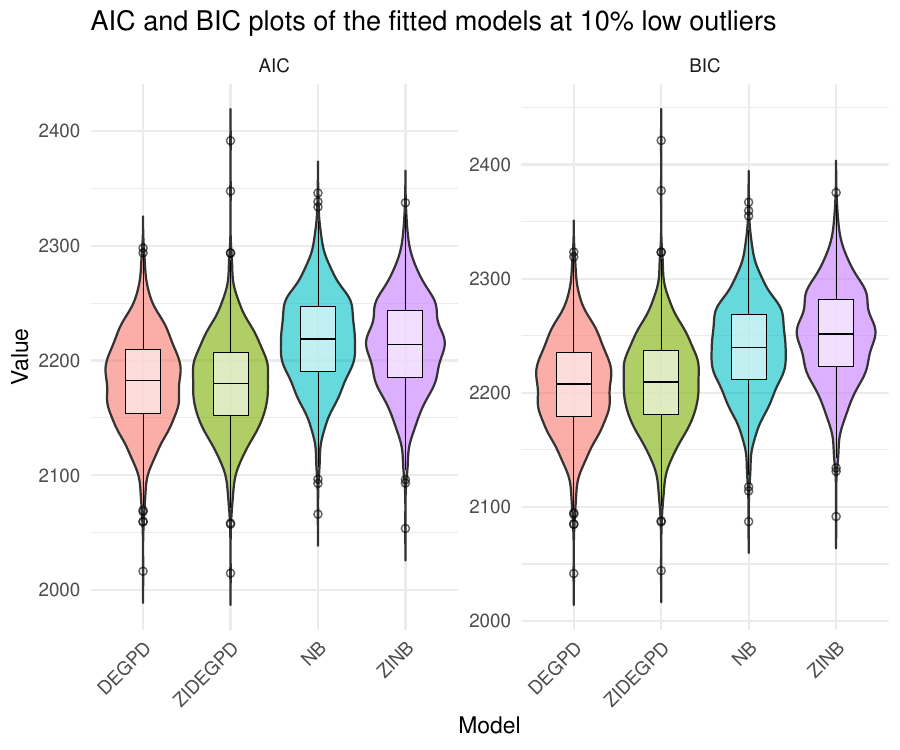}
\includegraphics[width=4.5cm,height=5.5cm]{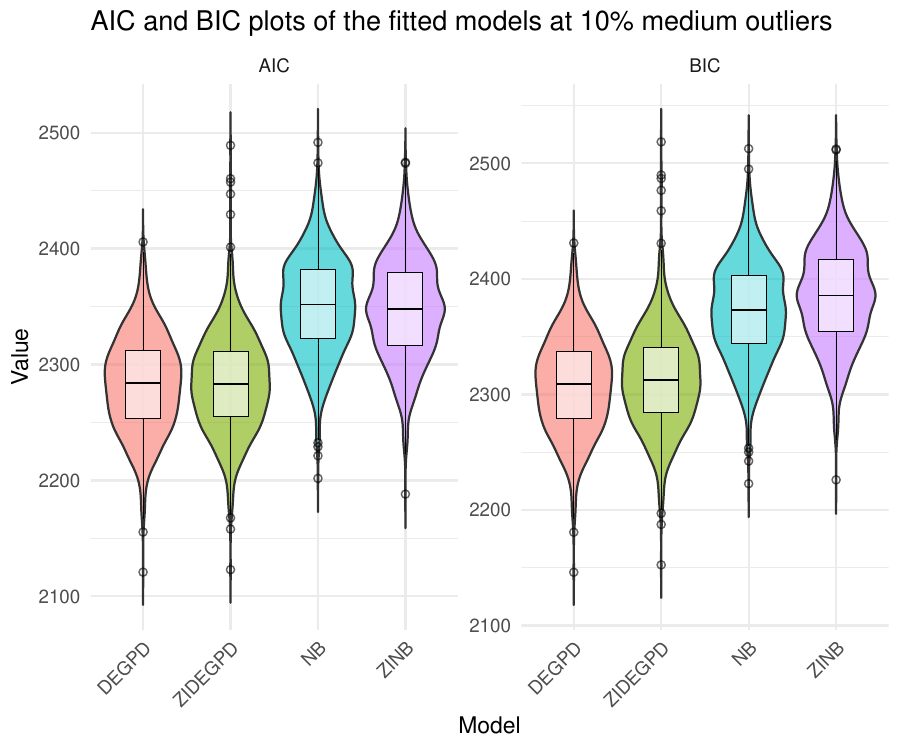}
\includegraphics[width=4.5cm,height=5.5cm]{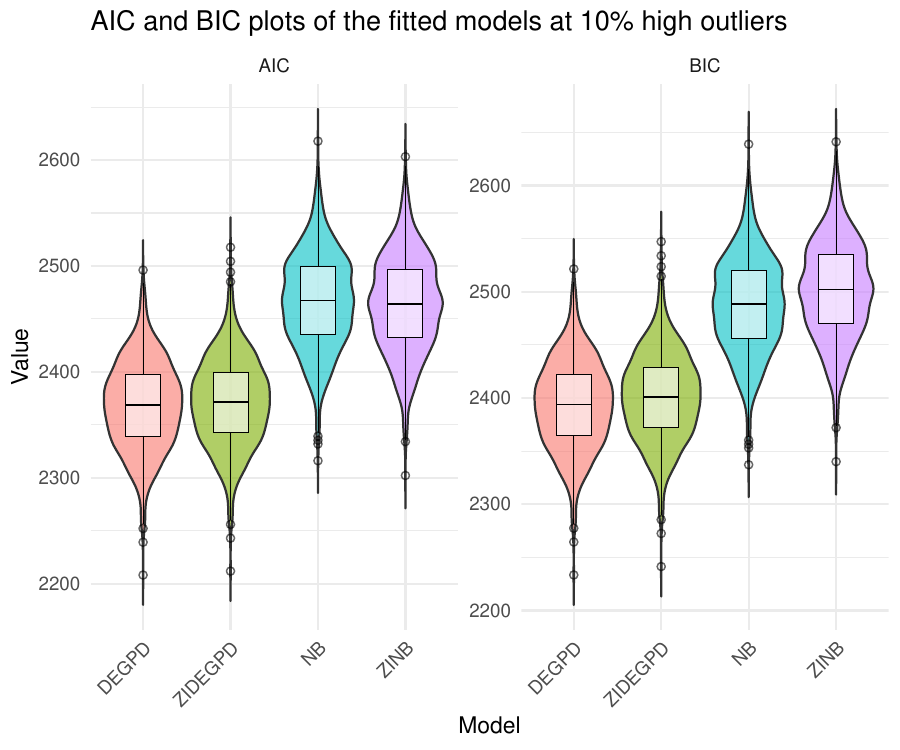}
\caption{Violin boxplots of AIC and BIC over $1000$ replications of the fitted models to data simulated from the negative binomial distribution with additional 25\% zeros and varying outlier proportions (1\%, 5\%, 10\%) in low, medium, and high levels.  }
\label{fig:enter-label}
\end{figure}

\begin{figure}
\centering
\includegraphics[width=4.5cm,height=5.5cm]{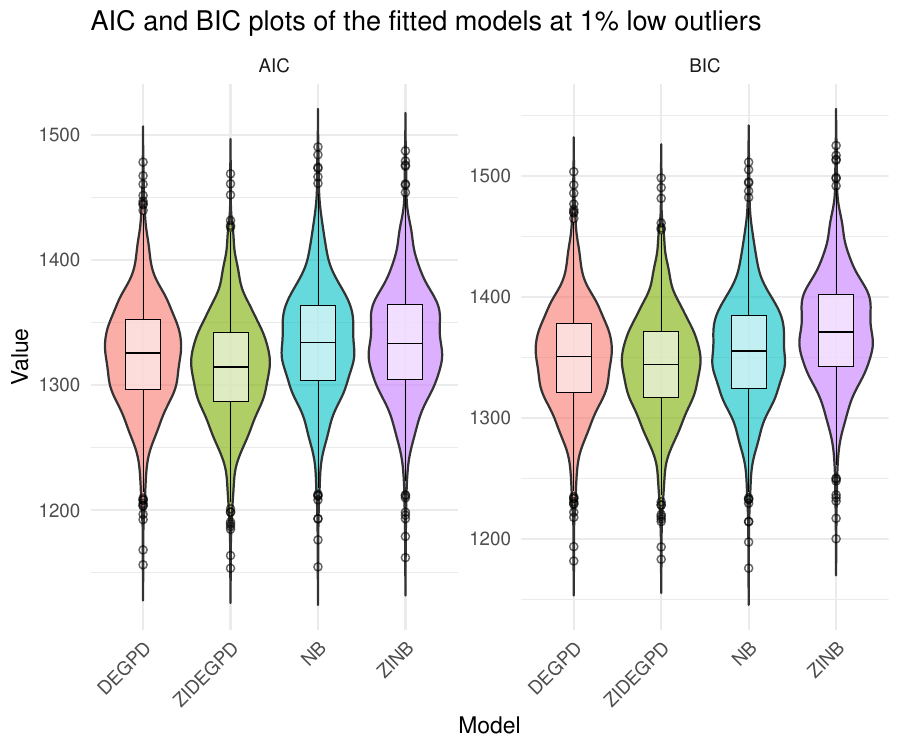}
\includegraphics[width=4.5cm,height=5.5cm]{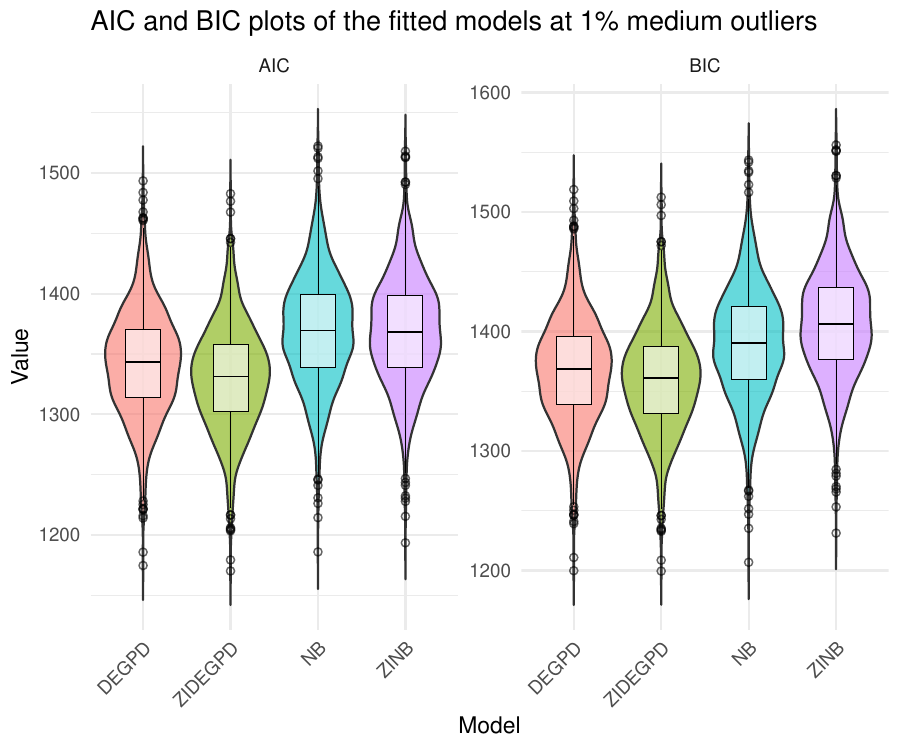}
\includegraphics[width=4.5cm,height=5.5cm]{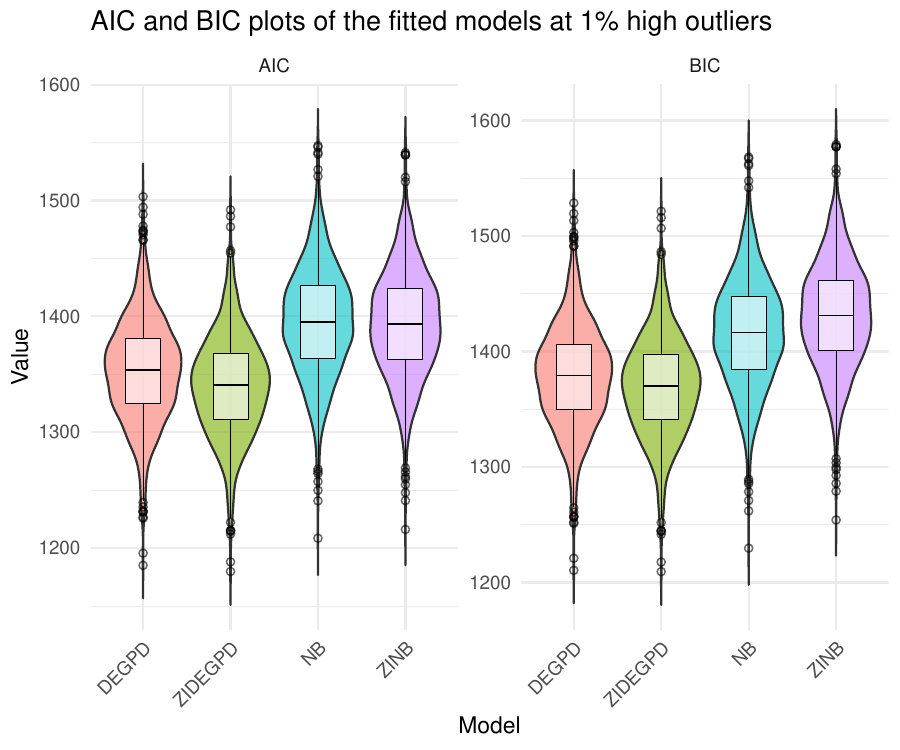}
\includegraphics[width=4.5cm,height=5.5cm]{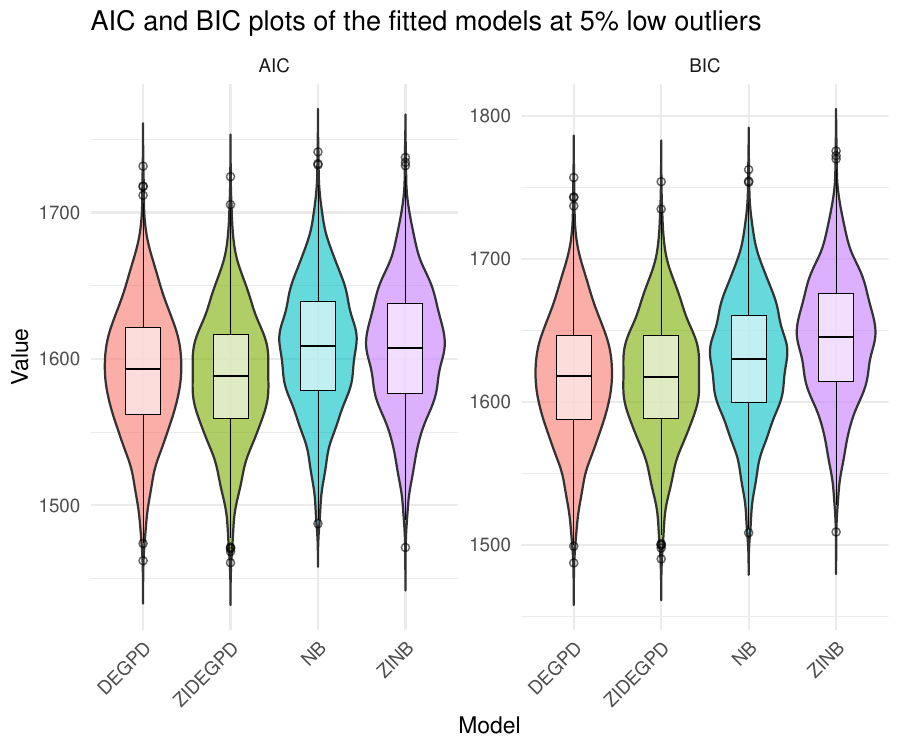}
\includegraphics[width=4.5cm,height=5.5cm]{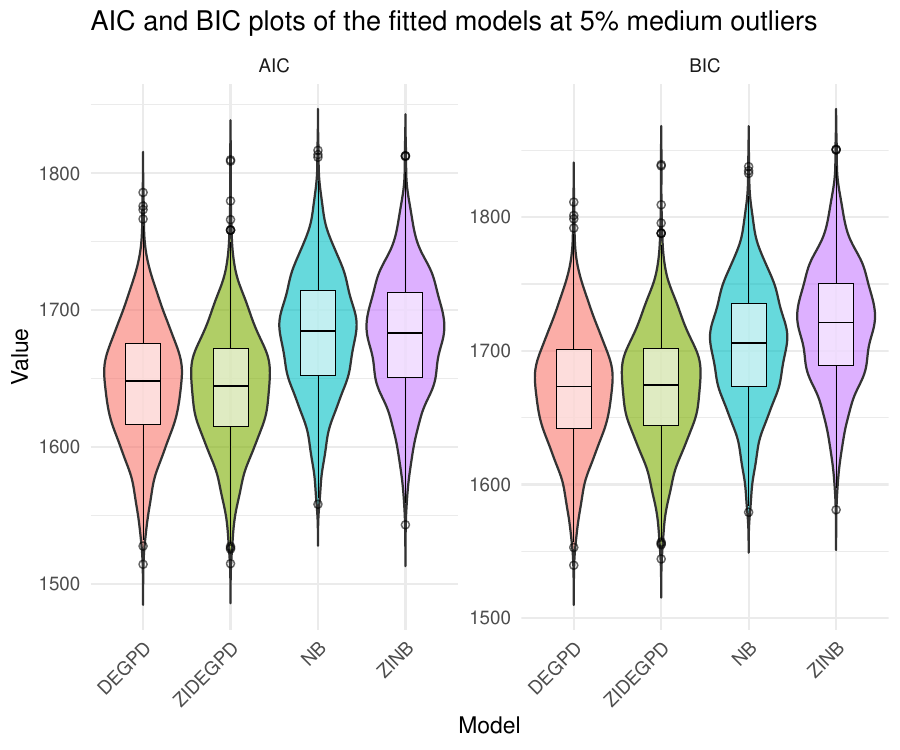}
\includegraphics[width=4.5cm,height=5.5cm]{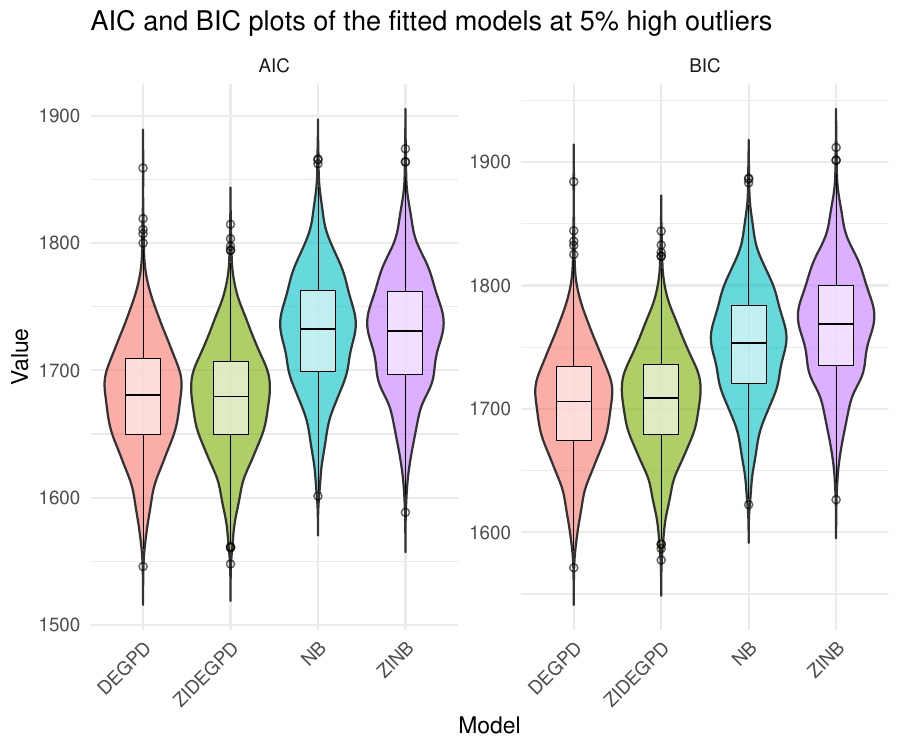}
\includegraphics[width=4.5cm,height=5.5cm]{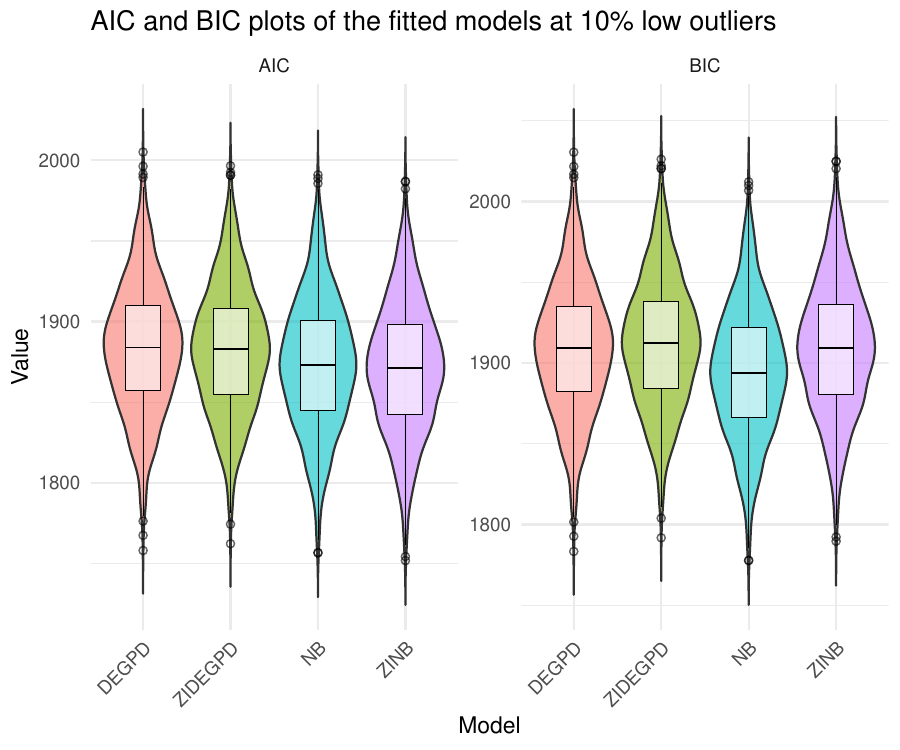}
\includegraphics[width=4.5cm,height=5.5cm]{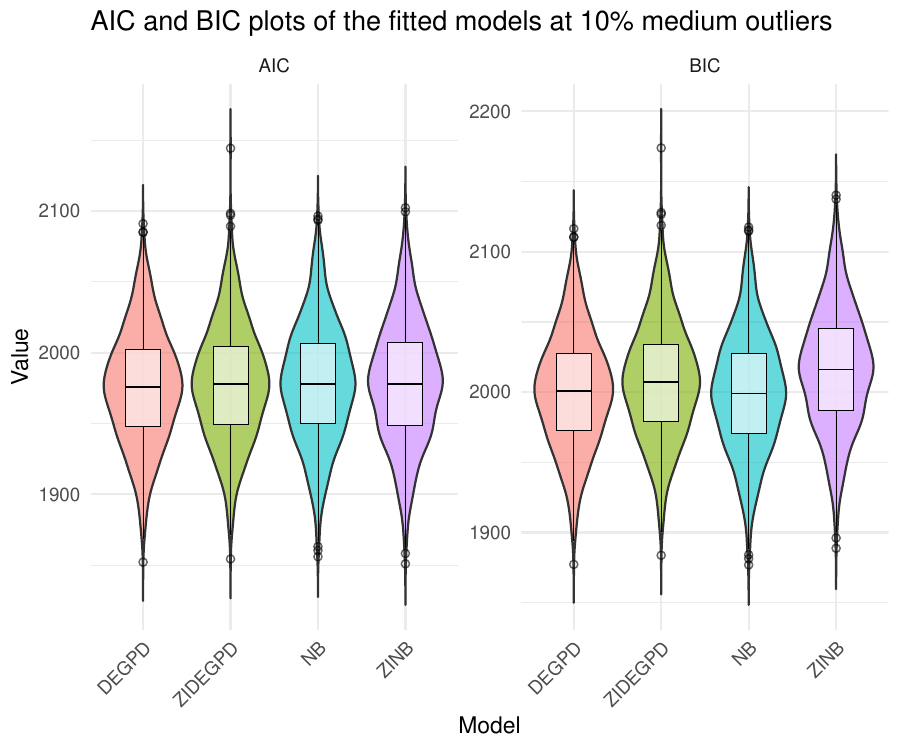}
\includegraphics[width=4.5cm,height=5.5cm]{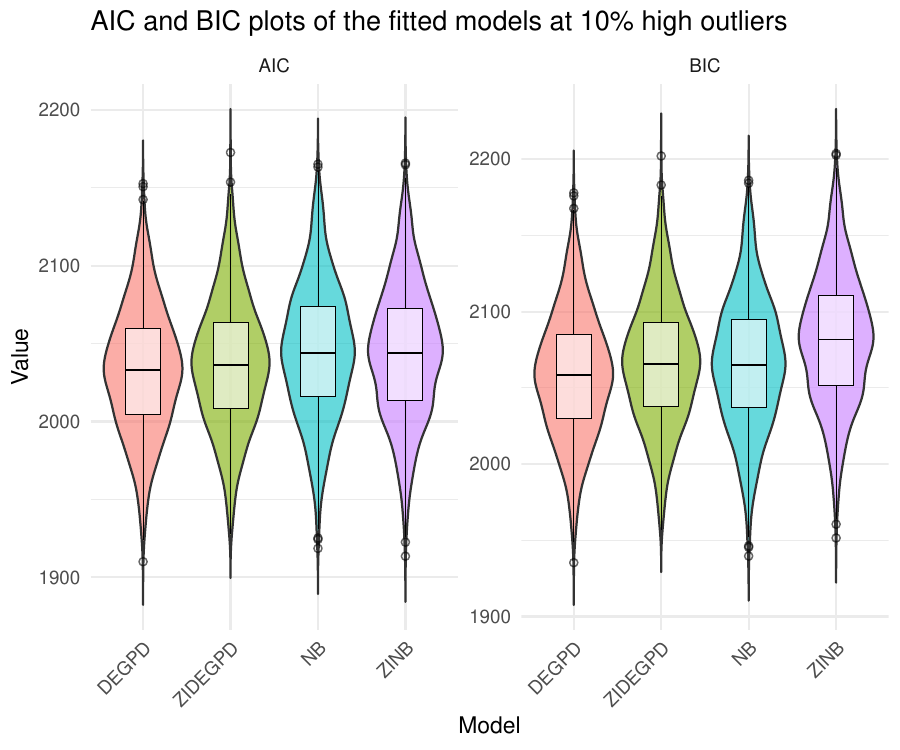}
\caption{Violin boxplots of AIC and BIC over $1000$ replications of the fitted models to data simulated from the negative binomial distribution with additional 50\% zeros and varying outlier proportions (1\%, 5\%, 10\%) in low, medium, and high levels.}
\label{fig:fitted-50}
\end{figure}

\begin{figure}
\centering
\includegraphics[width=4.5cm,height=5.5cm]{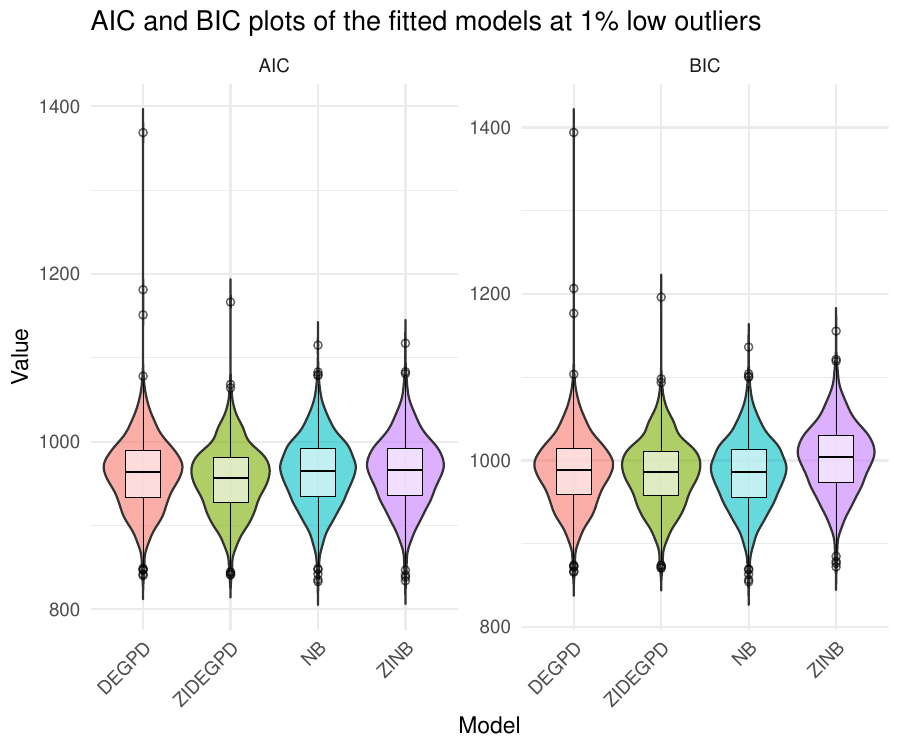}
\includegraphics[width=4.5cm,height=5.5cm]{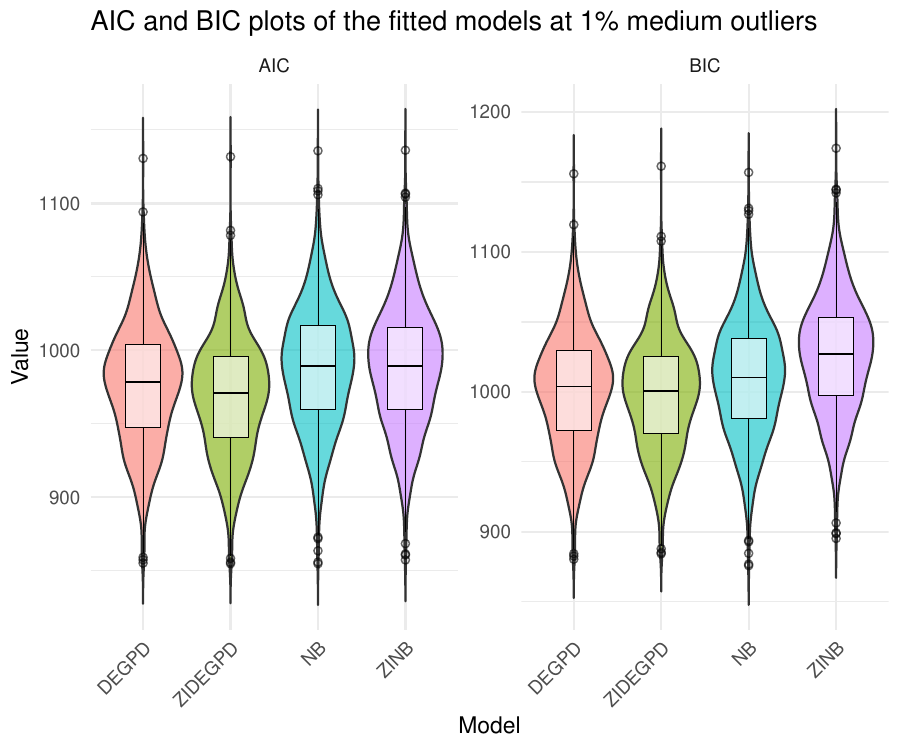}
\includegraphics[width=4.5cm,height=5.5cm]{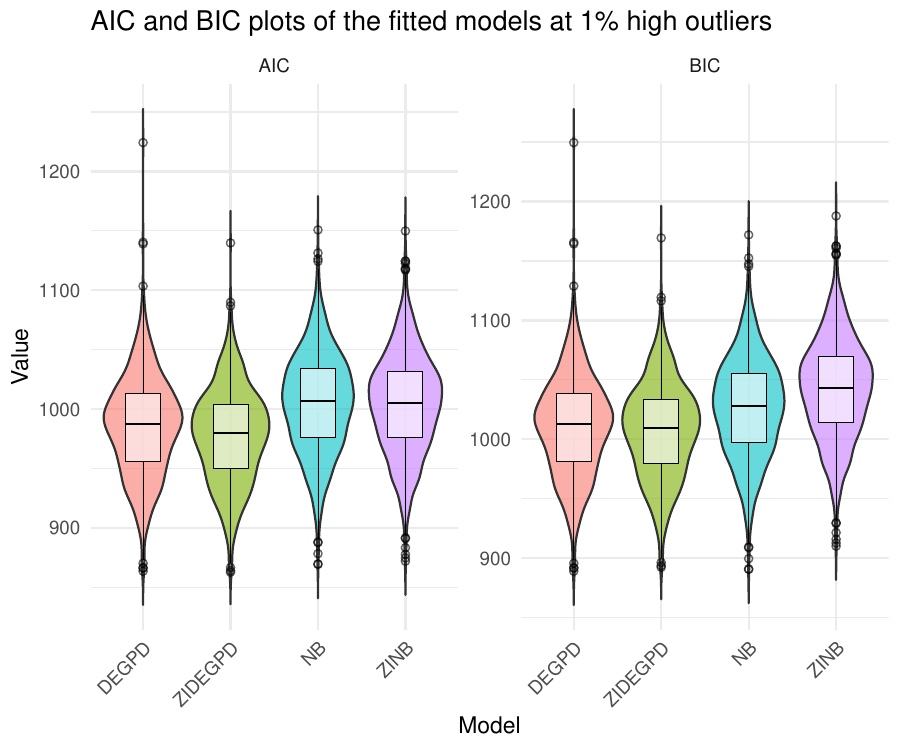}
\includegraphics[width=4.5cm,height=5.5cm]{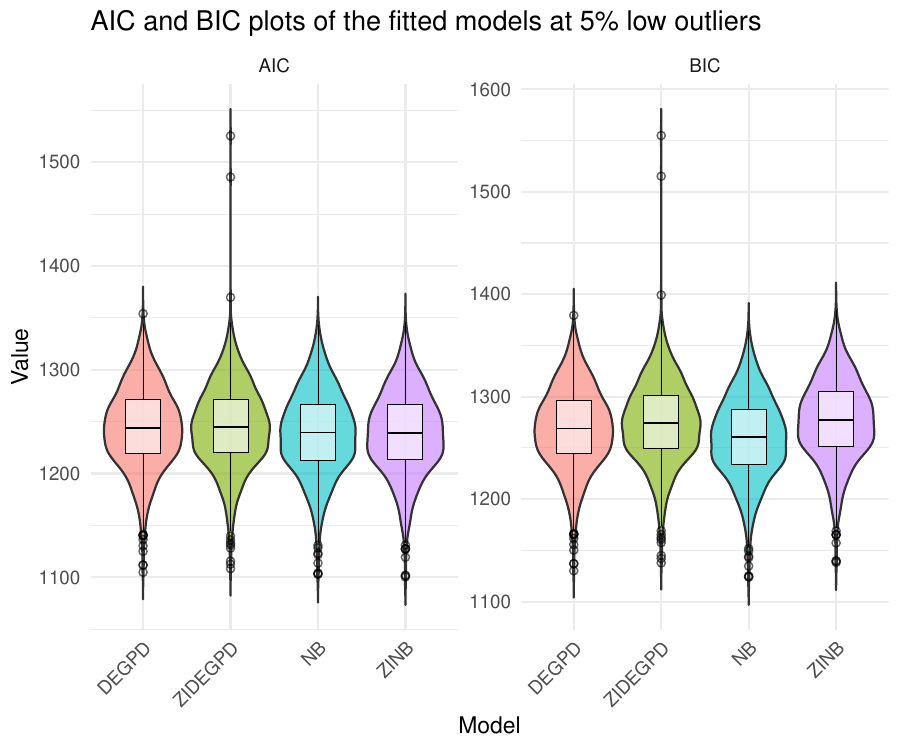}
\includegraphics[width=4.5cm,height=5.5cm]{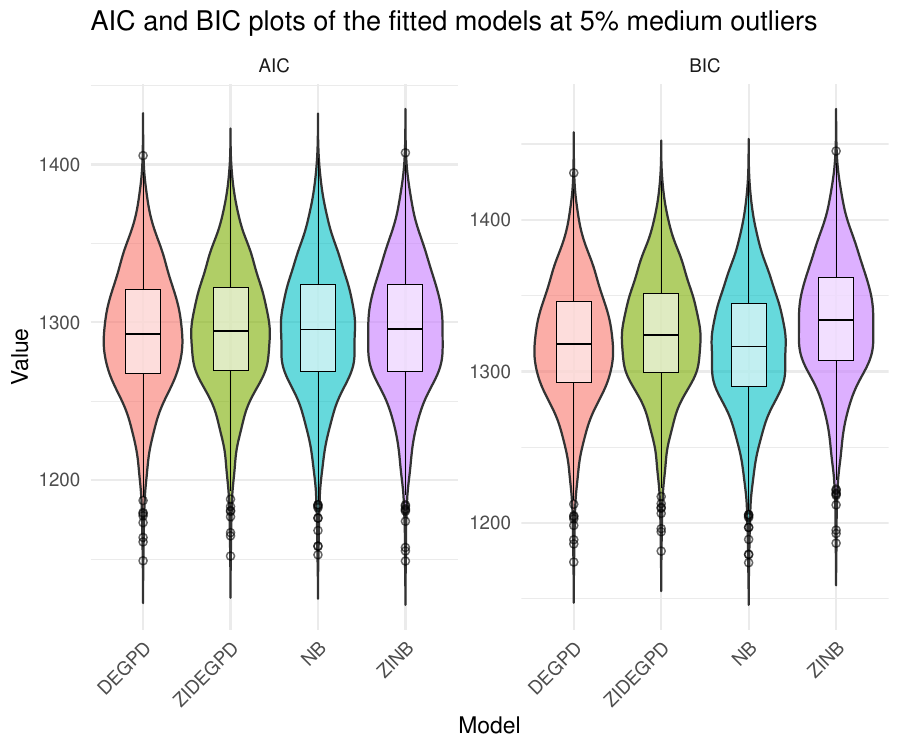}
\includegraphics[width=4.5cm,height=5.5cm]{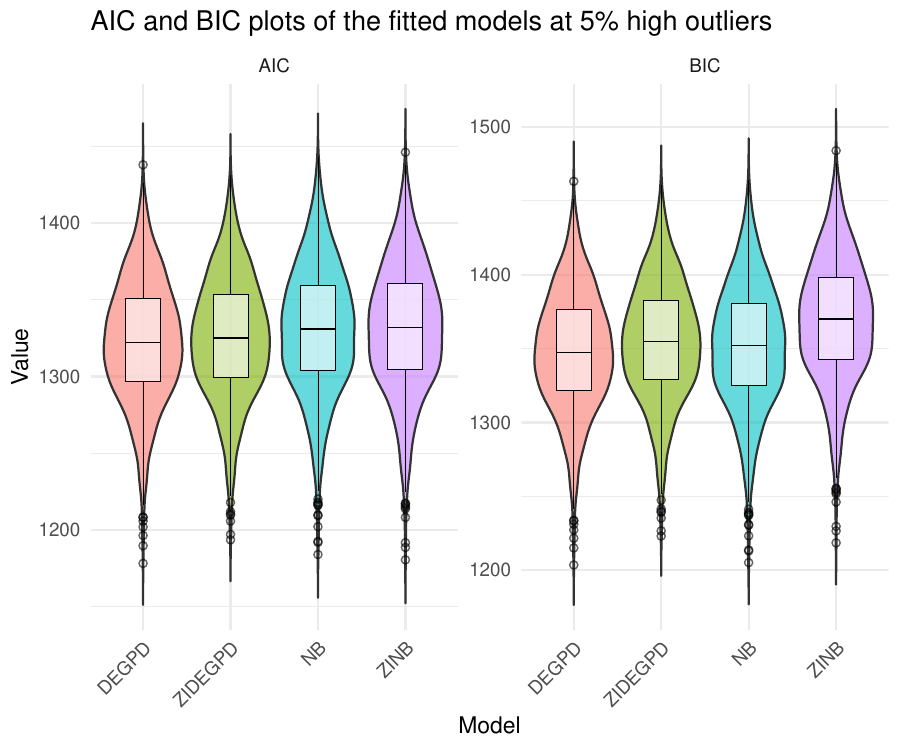}
\includegraphics[width=4.5cm,height=5.5cm]{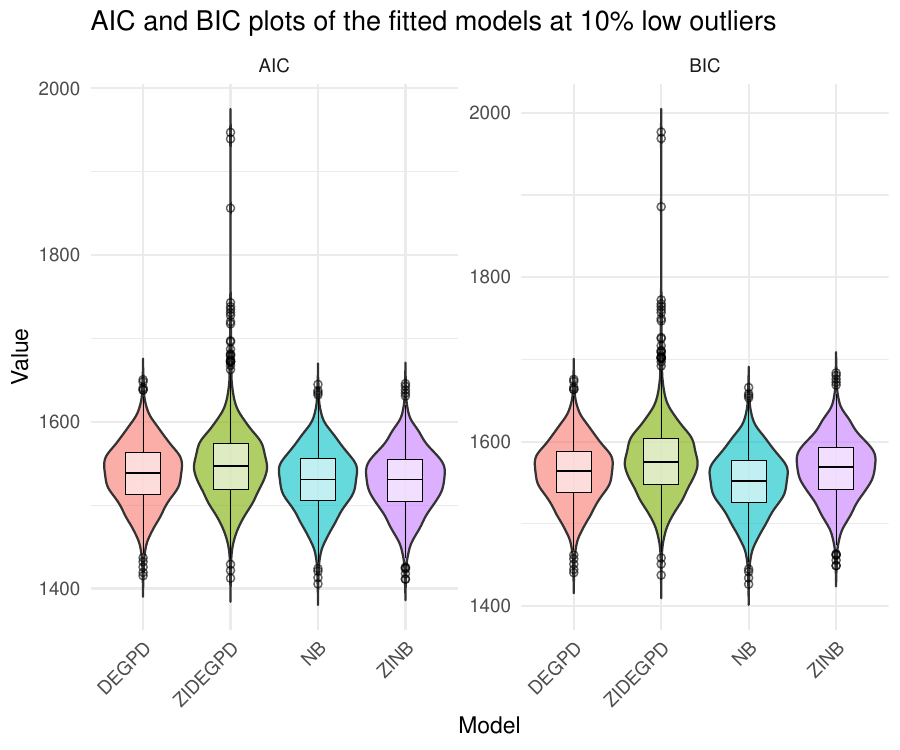}
\includegraphics[width=4.5cm,height=5.5cm]{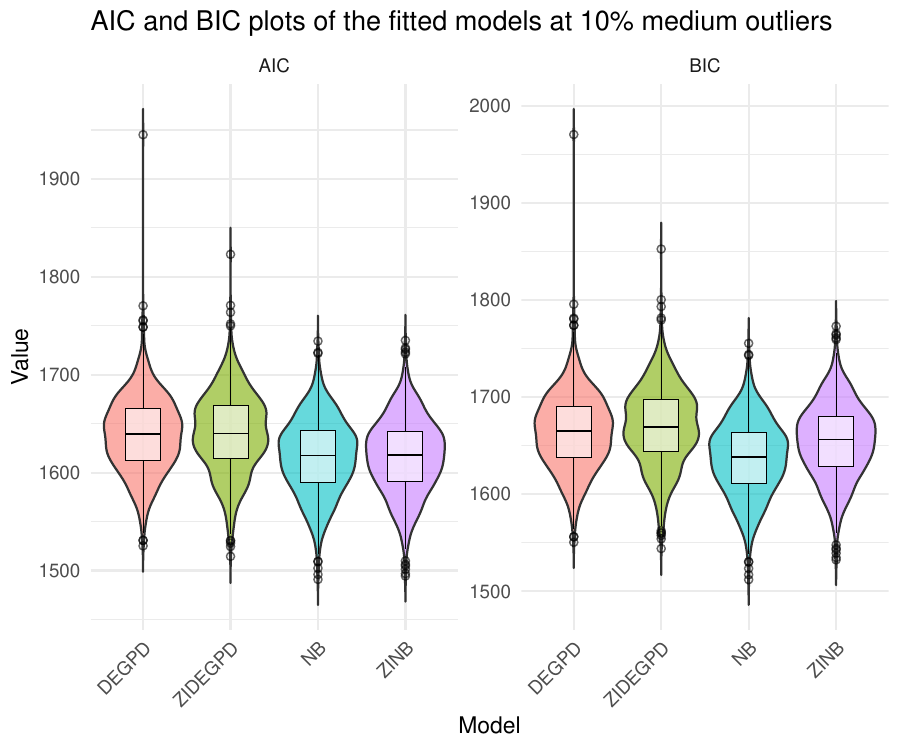}
\includegraphics[width=4.5cm,height=5.5cm]{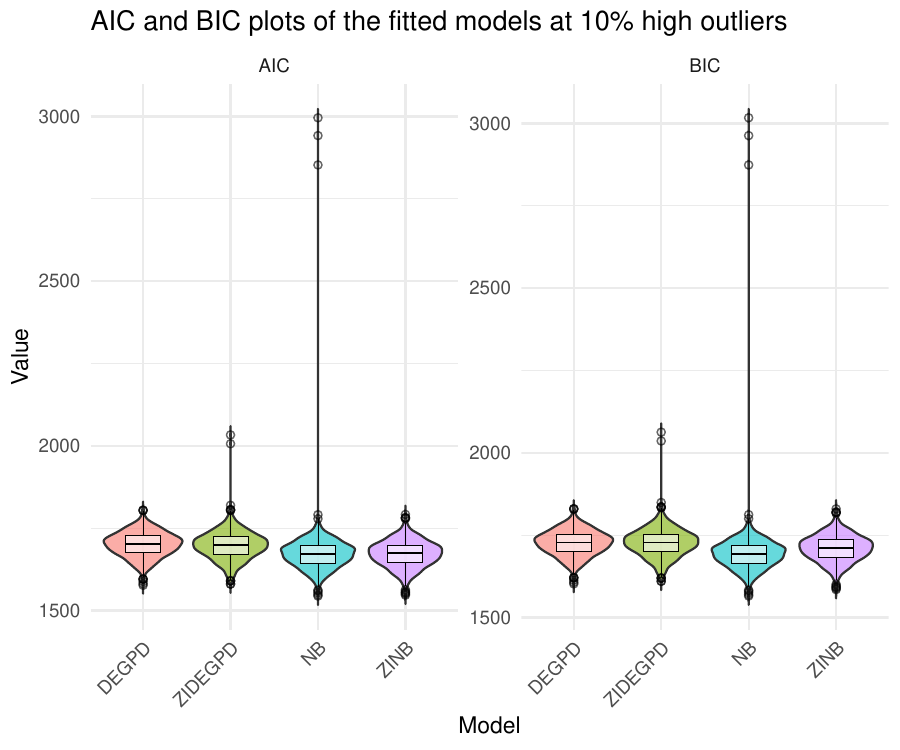}
\caption{Violin boxplots of AIC and BIC over $1000$ replications of the fitted models to data simulated from the negative binomial distribution with additional 70\% zeros and varying outlier proportions (1\%, 5\%, 10\%) in low, medium, and high levels.}
\label{fig:fitted-70}
\end{figure}

Figures~\ref{fig:no-zero-inflation}–\ref{fig:fitted-70} summarize the outcomes of all simulation scenarios, using violin boxplots to display the distribution of AIC and BIC values across 1,000 replications. In the baseline scenario, with no zero-inflation or outliers and data generated from a standard negative binomial distribution (NBD), both the NB and ZINB models performed well, achieving the lowest median AIC/BIC values. The DEGPD models, although more flexible, showed slightly higher AIC/BIC values due to the additional parameters addressing issues absent in the data. Results with introducing 25\% zero-inflation and outliers highlighted the superior performance of the DEGPD and ZIDEGPD models relative to NB and ZINB. As the proportion and magnitude of outliers increased, the ZIDEGPD and DEGPD models increasingly outperformed ZINB model. While ZINB effectively handles excess zeros, it relies on the NBD, which poorly captures heavy tails. While ZIDEGPD and DEGPD tail based on the tail index of generalized Pareto distribution, which is specifically designed to model extreme values, provide a better fit in the tail of the distribution.

At elevated levels of zero-inflation (50\% and 70\%), the application of a zero-inflated model became imperative. ZINB and ZIDEGPD models considerably surpassed the NB and DEGPD models. A significant discovery was that ZIDEGPD sustained, and in several instances enhanced, its performance superiority over ZINB as the degree of outlier contamination escalated. Despite comprising 70\% zeros, the tail behavior influenced by outliers continued to be a significant difference. ZIDEGPD's ability to simultaneously model excess zeros (through the zero-inflation component) and extreme counts (via the DEGPD count component) resulted in consistently lower and less variable AIC/BIC values, particularly under medium and high outlier conditions.

In summary, the simulations confirmed that in the presence of excess zeros, zero-inflated models (ZINB, ZIDEGPD) are universally superior. More importantly, the proposed DEGPD and ZIDEGPD models demonstrated superior robustness to outliers. Their performance gap over NB and ZINB widened as the frequency and magnitude of outliers increased. Overall, the ZIDEGPD model proved to be the most robust and well-performing model across the vast majority of scenarios, particularly in real-world conditions where both zero-inflation and outliers are present. It successfully addresses the core limitation of ZINB identified in the introduction—the failure to accurately model the tail of the distribution.

\section{Real applications}\label{sec:realapp}
\begin{figure}
\centering
\includegraphics[width=0.45\linewidth]{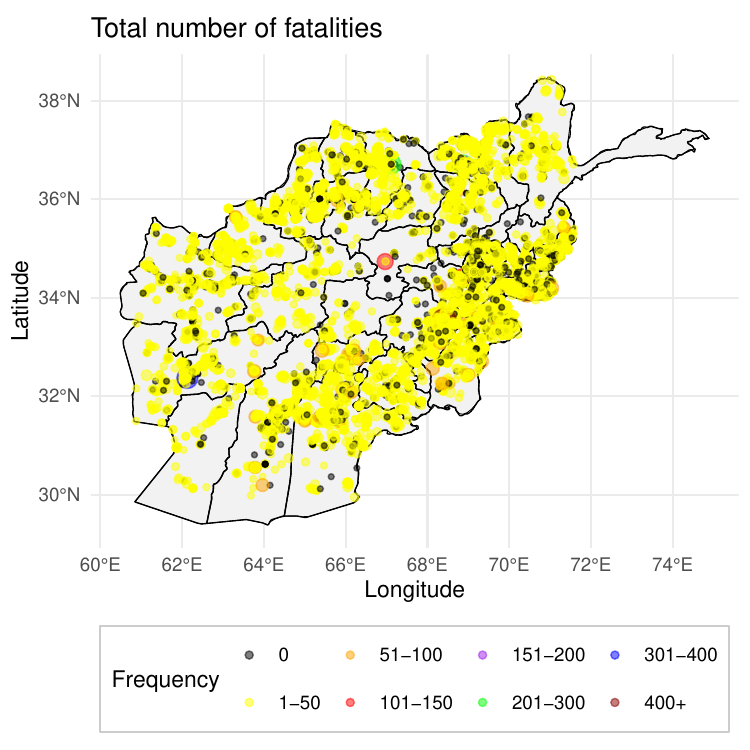}
\includegraphics[width=0.45\linewidth]{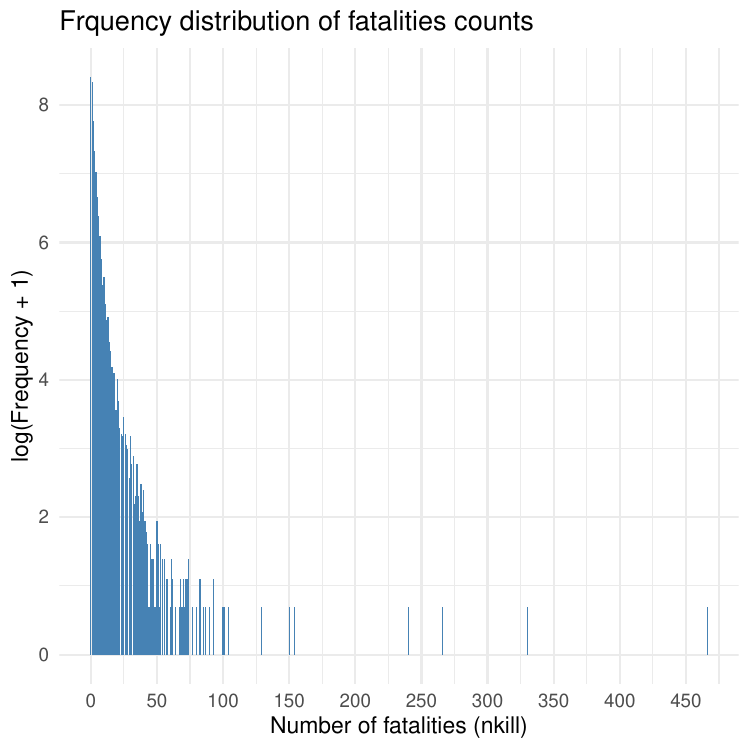}
\caption{\textit{Left:} Spatial distribution of total number of fatalities. \textit{Right:} Frequency distribution of fatality counts.}
\label{fig:freq-real}
\end{figure}

To demonstrate the performance of the models considered in the simulation study, we apply them to terrorism data from Afghanistan obtained from the Global Terrorism Database (GTD) \citep{GTD2022}. In this application, we model the total number of confirmed fatalities (nkill), which includes all victims and attackers who died as a direct result of each incident. The dataset covers terrorism incidents in Afghanistan from 1973 to 2020. For incidents with no reported fatalities, the count is recorded as zero. Geographic coordinates are provided for each incident. After removing missing values, 17,577 incidents with complete details remain. Of these, 25.45\% report no fatalities. Figure~\ref{fig:freq-real}(left) illustrates the spatial distribution of fatalities, where black points denote incidents with zero fatalities. Figure~\ref{fig:freq-real}(right) displays the frequency distribution of fatalities, where counts above 100 appear as outliers. In the regression setup, we use coordinates and their interaction as covariates.
Table~\ref{tab:model_comparison} presents the comparative results for all four fitted models. Across all models, the intercept is positive and highly significant, reflecting a strong baseline level of terrorism-related fatalities in Afghanistan. The coefficient for longitude is negative and statistically significant, suggesting that fatalities decrease towards the eastern regions. Similarly, latitude is consistently negative and significant, indicating higher fatality counts in southern provinces. The positive and significant interaction term (longitude:latitude) highlights spatial clusters where longitude and latitude jointly amplify fatality intensity.

In the DEGPD and ZIDEGPD models, both $\log(\gamma)$ and $\log(\alpha)$ are negative and highly significant, confirming strong overdispersion and heavy-tailed behavior. The $\text{logit}(\omega)$ parameter in ZIDEGPD is negative but statistically insignificant, implying that DEGPD adequately captures excess zeros without additional zero inflation. Among all models, DEGPD yields the lowest AIC and BIC values, indicating the best overall fit. These results suggest that fatalities in Afghanistan's terrorism data exhibit spatial dependence, overdispersion, and heavy-tail characteristics, best captured by the DEGPD family. Traditional models (NB and ZINB) fail to fully capture high-fatality outliers in certain regions. Spatial dependence should be further investigated using a spatial modeling framework.

\begin{figure}[h]
\centering
\includegraphics[width=0.23\linewidth]{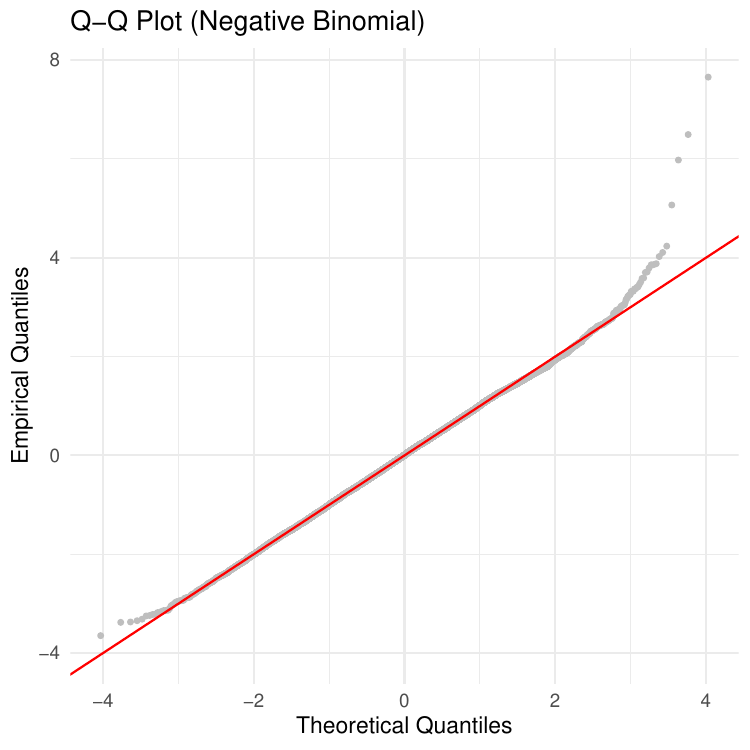}
\includegraphics[width=0.23\linewidth]{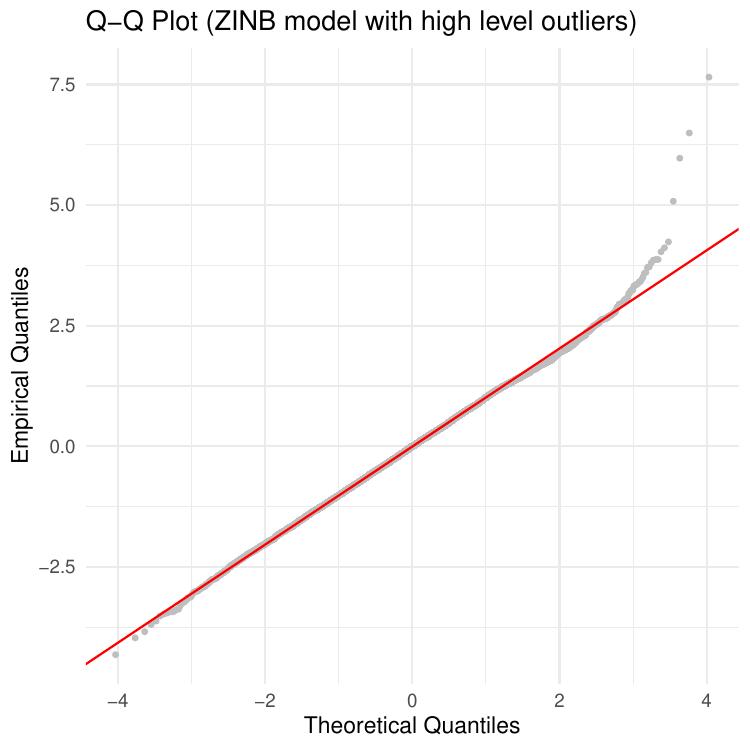}
\includegraphics[width=0.23\linewidth]{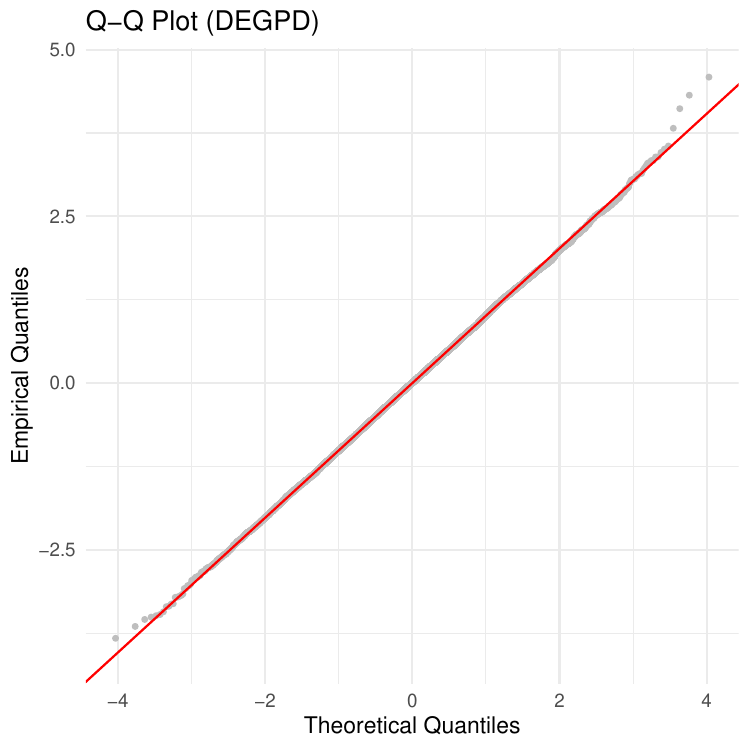}
\includegraphics[width=0.23\linewidth]{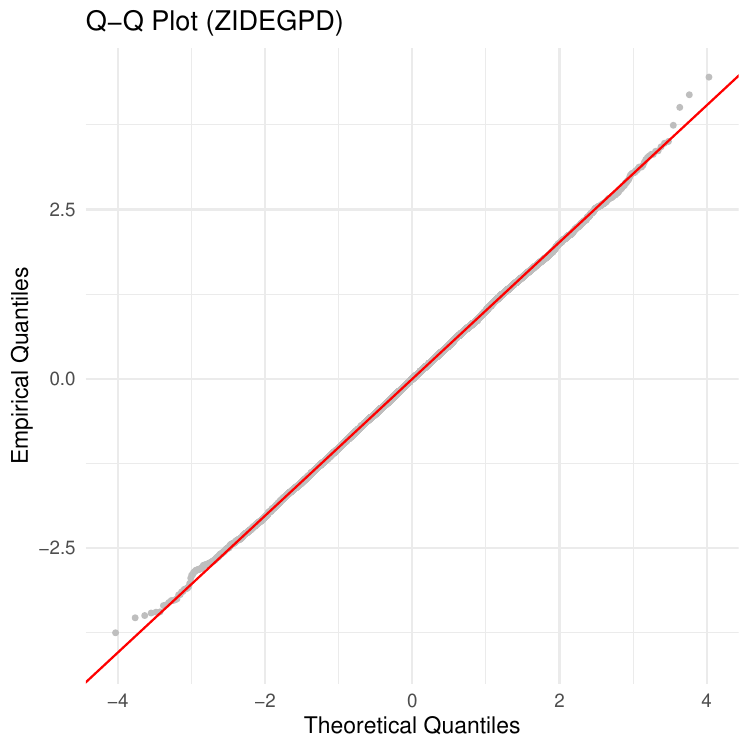}
\caption{Randomized residual QQ-plots of fitted models.}
\label{fig:qq-real}
\end{figure}

As shown in Figure~\ref{fig:qq-real}, randomized residual QQ-plots reveal that NB and ZINB models underestimate the impact of high-fatality incidents, producing deviations in the upper tails. In contrast, DEGPD and ZIDEGPD fit well across the entire distribution, including both tails, effectively capturing extreme fatality incidents or outliers.

\begin{table}[htbp]
\centering
\caption{Comparison of overdispersed and heavy-tailed models for terrorism fatalities.}
\label{tab:model_comparison}
\begin{tabular}{lcccc}
\toprule
 & NegBin & ZINB & DEGPD & ZI-DEGPD \\
\midrule
(Intercept) & 31.974* & 31.979*** & 33.420** & 32.230** \\
longitude & -0.468* & -0.468*** & -0.480** & -0.470** \\
latitude & -0.879* & -0.879*** & -0.880** & -0.840** \\
longitude:latitude & 0.0130* & 0.0130*** & 0.010** & 0.010** \\
$\log(\gamma)$ & – & – & -1.350*** & -1.220*** \\
$\log(\alpha)$ & – & – & -1.660*** & -1.100*** \\
$\text{logit}(\omega)$ & – & -15.790*** & – & -0.730 \\
\midrule
AIC & 46787.77 & 46789.77 & \textbf{46687.09} & 46688.20 \\
BIC & 46826.75 & 46836.55 & \textbf{46733.87} & 46742.78 \\
\bottomrule
\end{tabular}
\smallskip
\parbox{\textwidth}{\raggedright Significance codes: *** $p < 0.001$; ** $p < 0.01$; * $p < 0.05$}
\end{table}

\section{Conclusion}
\label{sec:conl}
In this study, we have proposed a modeling framework (DEGPD and ZIDEGPD) that exhibits heavy-tailed behavior and flexible bulk-part properties to address the challenges of count data with varying levels of zeros and outliers. Unlike traditional models such as NB or ZINB, the proposed heavy-tailed models simultaneously accommodate overdispersion, zero-inflation, and data heterogeneity within a unified framework. The bulk and tail parameters of these models (Sections~\ref{sec:degpd} and \ref{sec:zidegpd}) allow them to adapt effectively to diverse empirical scenarios where standard methods fail to provide satisfactory fits.

Simulation studies confirmed the efficiency of the heavy-tailed count data models across most scenarios, even with high proportions of zeros. Real-world applications demonstrated that DEGPD and ZIDEGPD outperform conventional models in terms of goodness-of-fit. The ability to handle outliers and high proportions of zeros makes DEGPD and ZIDEGPD particularly valuable for practical applications in health, insurance, ecology, and social sciences, where atypical counts are frequently observed. Future research may extend this work toward nonlinear regression models and Bayesian estimation frameworks.

\appendix
\section{Proof of Theorem~\ref{theorem1}}\label{theorem-proof}
\begin{proof}
For any integer $k \in \mathbb{N}_0$,
\[
\{|X| \geq k\} \stackrel{d}{=} \{X \geq k\},
\]
since $|X| \geq k$ if and only if $X \geq k$. Therefore,
\[
S_Z(k) = \Pr(Z \geq k) = \Pr(X \geq k) = S_X(k).
\]
By definition, the survival function of the continuous EGPD is:
\[
S_X(k) = 1 - F_X(k) = 1 - \left[ 1 - \left(1 + \frac{k \gamma}{\sigma} \right)^{-1/\gamma} \right]^\alpha.
\]
Define
\[
a_k := \left(1 + \frac{k \gamma}{\sigma} \right)^{-1/\gamma}.
\]
As $k \to \infty$, $a_k \to 0$ and
\[
a_k \sim \left( \frac{k \gamma}{\sigma} \right)^{-1/\gamma} = \left( \frac{\sigma}{\gamma} \right)^{1/\gamma} k^{-1/\gamma}.
\]
Using the first-order expansion $1 - (1 - a)^\alpha \sim \alpha a$ for small $a$, we obtain
\[
S_X(k) = 1 - (1 - a_k)^\alpha \sim \alpha a_k \sim \alpha \left( \frac{\sigma}{\gamma} \right)^{1/\gamma} k^{-1/\gamma}, \quad (k \to \infty).
\]
From this result for the continuous EGPD, we derive for the DEGPD:
\[
S_Z(k) \sim C k^{-1/\gamma}, \quad \text{where } C = \alpha \left( \frac{\sigma}{\gamma} \right)^{1/\gamma} > 0.
\]
Hence, $S_Z \in RV_{-1/\gamma}$, meaning the discrete survival function is regularly varying with index $-1/\gamma$. This satisfies the definition of regular variation in \citet{hitz_davis_samorodnitsky_2024} for discrete distribution pertaining a heavy tail. Consequently, the DEGPD preserves the same tail index as its continuous counterpart, and $F_Z$ belongs to $\mathrm{d\text{-}MDA}_\gamma$.
\end{proof}

\section{Likelihood expressions}\label{lik-appendix}
We compare the four count regression models in this study. The log-likelihood expressions for all models are derived to align with the distributional regression framework. The log-likelihood of the NB model (Section~\ref{sec:negb}) is
\begin{equation}
\begin{aligned}
\mathcal{L}(\boldsymbol{\psi}) &= \sum_{i=1}^n \log \Pr(Z_i = k_i \mid \mu(\mathbf{x}_i), \alpha(\mathbf{x}_i)) \\
&= \sum_{i=1}^n \left\{ \log \Gamma(k_i + \alpha(\mathbf{x}_i)) - \log \Gamma(\alpha(\mathbf{x}_i)) - \log(k_i!) \right. \\
&\quad + \left. \alpha(\mathbf{x}_i) \log \left( \frac{\alpha(\mathbf{x}_i)}{\alpha(\mathbf{x}_i) + \mu(\mathbf{x}_i)} \right) + k_i \log \left( \frac{\mu(\mathbf{x}_i)}{\alpha(\mathbf{x}_i) + \mu(\mathbf{x}_i)} \right) \right\}.
\end{aligned}
\label{eq:app_nb_loglik}
\end{equation}
The log-likelihood of the ZINB model (Section~\ref{sec:zinegb}) is
\begin{equation}
\begin{aligned}
\mathcal{L}(\boldsymbol{\psi}) &= \sum_{i=1}^n I_{\{0\}}(k_i) \log \left[ \omega(\mathbf{x}_i) + (1 - \omega(\mathbf{x}_i)) \left( \frac{\alpha(\mathbf{x}_i)}{\alpha(\mathbf{x}_i) + \mu(\mathbf{x}_i)} \right)^{\alpha(\mathbf{x}_i)} \right] \\
&\quad + \sum_{i=1}^n (1 - I_{\{0\}}(k_i)) \log \left\{ (1 - \omega(\mathbf{x}_i)) \frac{\Gamma(k_i + \alpha(\mathbf{x}_i))}{\Gamma(\alpha(\mathbf{x}_i)) k_i!} \right. \\
&\quad \times \left. \left( \frac{\alpha(\mathbf{x}_i)}{\alpha(\mathbf{x}_i) + \mu(\mathbf{x}_i)} \right)^{\alpha(\mathbf{x}_i)} \left( \frac{\mu(\mathbf{x}_i)}{\alpha(\mathbf{x}_i) + \mu(\mathbf{x}_i)} \right)^{k_i} \right\}.
\end{aligned}
\label{eq:app_zinb_loglik}
\end{equation}
The log-likelihood of the DEGPD model (Section~\ref{sec:degpd}) is
\begin{equation}
\begin{aligned}
\mathcal{L}(\boldsymbol{\psi}) &= \sum_{i=1}^n \log \left[ \mathcal{R}(\mathcal{H}(k_i + 1; \sigma(\mathbf{x}_i), \gamma(\mathbf{x}_i)); \alpha(\mathbf{x}_i)) \right. \\
&\quad - \left. \mathcal{R}(\mathcal{H}(k_i; \sigma(\mathbf{x}_i), \gamma(\mathbf{x}_i)); \alpha(\mathbf{x}_i)) \right] \\
&= \sum_{i=1}^n \log \left[ \left( 1 - \left( 1 + \frac{(k_i + 1)\gamma(\mathbf{x}_i)}{\sigma(\mathbf{x}_i)} \right)^{-1/\gamma(\mathbf{x}_i)} \right)^{\alpha(\mathbf{x}_i)} \right. \\
&\quad - \left. \left( 1 - \left( 1 + \frac{k_i \gamma(\mathbf{x}_i)}{\sigma(\mathbf{x}_i)} \right)^{-1/\gamma(\mathbf{x}_i)} \right)^{\alpha(\mathbf{x}_i)} \right].
\end{aligned}
\label{eq:app_degpd_loglik}
\end{equation}
The log-likelihood of the ZIDEGPD model (Section~\ref{sec:zidegpd}) is
\begin{equation}
\begin{aligned}
\mathcal{L}(\boldsymbol{\psi}) &= \sum_{i=1}^n I_{\{0\}}(k_i) \log \left[ \omega(\mathbf{x}_i) + (1 - \omega(\mathbf{x}_i)) \mathcal{R}(\mathcal{H}(1; \sigma(\mathbf{x}_i), \gamma(\mathbf{x}_i)); \alpha(\mathbf{x}_i)) \right] \\
&\quad + \sum_{i=1}^n (1 - I_{\{0\}}(k_i)) \log(1 - \omega(\mathbf{x}_i)) \\
&\quad \times \left[ \mathcal{R}(\mathcal{H}(k_i + 1; \sigma(\mathbf{x}_i), \gamma(\mathbf{x}_i)); \alpha(\mathbf{x}_i)) \right. \\
&\quad - \left. \mathcal{R}(\mathcal{H}(k_i; \sigma(\mathbf{x}_i), \gamma(\mathbf{x}_i)); \alpha(\mathbf{x}_i)) \right],
\end{aligned}
\label{eq:app_zidegpd_loglik}
\end{equation}
where $I_A(\cdot)$ is the indicator function for set $A$. The link functions for the model parameters are discussed in Section~\ref{distregression}. Model fitting is performed by maximizing the log-likelihood function $\mathcal{L}(\boldsymbol{\psi})$. The implementation also accommodates non-parametric components if needed.

\section*{Acknowledgments}
We acknowledge the role of large language models in assisting with language correction.

\section*{Disclosure statement}
The authors declare that they have no conflict of interest.

\section*{Funding}
This study did not receive any specific funding.

\section*{Notes on contributors}
Both authors contributed equally to this research.

\section*{Data and code availability}
The source of the data used in this study is given in the paper and codes are made available at \url{https://github.com/touqeerahmadunipd/FlexibleCountModel}.
\bibliographystyle{apalike}
\bibliography{bibliography}

\end{document}